\documentclass{article}
\usepackage{pgfplots}
\usepackage{lipsum}
\usepackage{caption}
\usepackage{subcaption}
\usepackage{authblk}
\usepackage[english]{babel}
\usepackage{amsfonts}
\usepackage{amsmath}
\usepackage[top=2cm, bottom=2cm, left=2cm, right=2cm]{geometry}
\usepackage{fancyhdr}
\usepackage{multicol}
\usepackage{bbold}
\usepackage{epsfig}
\usepackage{graphics}
\usepackage{epsfig}
\usepackage{lipsum}
\usepackage{fancyhdr}
\usepackage{booktabs}

\usepackage{wrapfig}
\usepackage[export]{adjustbox}
\usepackage{rotating}
\usepackage{amsmath}

\usepackage{hyperref} 
\usepackage{cleveref}

\usepackage{cite}
\renewenvironment{abstract}{
	
	\hfill\begin{minipage}{0.95\textwidth}
		\rule{\textwidth}{1pt}}
	{\par\noindent\rule{\textwidth}{1pt}\end{minipage}}

\makeatletter
\usepackage{orcidlink}
\makeatother
\begin{document}
	\title{\textbf{Parity Symmetry Breaking of spin-$j$ coherent state superpositions in Gaussian noise channel}}
	\author[1]{\textbf{B. El Alaoui \normalsize\orcidlink{0009-0008-4261-7961}}}
	\author[1,2]{\textbf{A. Slaoui \normalsize\orcidlink{0000-0002-5284-3240}}{\footnote {Corresponding author: {\sf abdallah.slaoui@um5s.net.ma}}}}
	\author[1,2]{\textbf{A. Lakhfif \normalsize\orcidlink{0000-0001-8004-4772}}}
	\author[1,2]{\textbf{ R. Ahl Laamara \normalsize\orcidlink{0000-0002-8254-9085}}}
	\affil[1]{\small LPHE-Modeling and Simulation, Faculty of Sciences, Mohammed V University in Rabat, Rabat, Morocco.}
	\affil[2]{\small Centre of Physics and Mathematics, CPM, Faculty of Sciences, Mohammed V University in Rabat, Rabat, Morocco.}
	%\affil[3]{\small College of Physical and Chemical Sciences, Hassan II Academy of Science and Technology, Rabat, Morocco.}
	%
	%
	\maketitle
 \begin{center}
  \textbf{Abstract}
	\end{center}
	\begin{abstract}
	The Wigner function and Wigner-Yanase skew information are connected through quantum coherence. States with high skew information often exhibit more pronounced negative regions in their Wigner functions, indicative of quantum interference and non-classical behavior. Thus, the relationship between these two concepts is that states with high quantum coherence tend to display more non-classical features in their Wigner functions. By exploiting this relationship, which manifests as parity symmetry and asymmetry, we analyze parity symmetry and asymmetry in the superposition of two spin coherent states for a spin-$1/2$, as well as for a general spin-$j$. This analysis shows that the preservation of the parity asymmetry, or the violation of the parity symmetry, correlates with an increase in the value of spin $j$. Additionally, we investigate the behavior of parity symmetry and asymmetry of these states subjected to a Gaussian noise channel. Specifically, we examine how this parity symmetry and asymmetry change and identify the points at which parity symmetry is violated in the spin-$1/2$ cat state. Notably, the violation of parity symmetry becomes more pronounced at higher values of the decoherence parameter $s$. Our study shows how the spin value $j$ affects the breaking of parity symmetry in general spin-$j$ cat states that are hit by Gaussian noise.

\end{abstract}
	
	\vspace{0.5cm}
	
	\section{Introduction}
	Quantifying the information content or uncertainty of a quantum state is a crucial issue in quantum information theory \cite{Frankel2019,Augenblick2021}. Within the framework of quantum mechanics, the uncertainty of a state essentially refers to the widely used von Neumann entropy \cite{vonNeumann}. This quantity, which stems from its deep links with information theory and statistical mechanics, describes the degree of mixing of a state with respect to its spectral decomposition. Despite its significance and operational meaning, it is already recognized that the von Neumann entropy has certain conceptual limitations in describing quantum systems exhibiting specific symmetries \cite{Balian1987}. In particular, when the quantum system possesses a conserved quantity, some observables can be more difficult to measure than others. It is then logical to consider an uncertainty measure that takes into account this restriction on the measurement of observables. Accordingly, different measures are suggested, such as the Wigner-Yanase skew information, which was introduced by Wigner and Yanase \cite{Wigner1963,Wigner1964}. In contrast to the entropic approach, the Wigner-Yanase skew information measures the information content of a quantum state skew to a conserved observable \cite{Shunlong2003,Shunlong2019,Shunlong2012}. It differs significantly from, yet maintains a deep connection with, the quantum entropies \cite{Kak2007, Chehade2019}. Besides its original meaning, some earlier studies show that the Wigner-Yanase skew information is closely related to many quantum effects such as uncertainty \cite{Chen2016, Fu2019, ZhangG2021,Abouelkhir2023,Xu2022}, non-commutativity \cite{Mohamed2017, Cheng2014}, information geometry \cite{Gibilisco2003,Hasegawa2003,Amghar2023,Hasegawa1997}, Bell-type inequalities \cite{Chen2005,Li12016,Biswas2023}, quantum correlations \cite{Slaoui2019,Dakir2023,Slaoui2023} and quantum coherence \cite{Frerot2016,Lei2016, Jafari2020,Li2016,Pires2020}. It is shown by Luo that it can also be regarded as a version of quantum Fisher information, which has crucial implications in the field of quantum estimation and quantum metrology \cite{Toth2014, Banik2017}. Another important fact is that skew information can be used to quantify the asymmetry of quantum states with respect to quantum channels, i.e., the degree to which a state does not commute with the generators of a specific symmetry group \cite{Sun2021,Pinto2021}.\par
	
	On the other hand, quantum decoherence stands as a fundamental phenomenon in quantum information theory, posing a significant challenge to the realization of robust quantum computations and communications \cite{Jiang2007,Slaouichap,KirdiIkken2023,Gheorghiu2015}. It arises from the intricate interaction of a quantum system with its external environment, causing the system's delicate quantum coherence to degrade over time. The coherence loss leads to the emergence of classical-like behavior, ultimately hindering the implementation of quantum algorithms and the faithful transmission of quantum information. Understanding and management of quantum decoherence play a pivotal role in advancing the field, with ongoing efforts focused on developing error-resistant quantum technologies and algorithms. Noisy Gaussian channels represent a class of quantum communication channels that introduce continuous-variable quantum systems to the realm of information transmission. These channels are characterized by the presence of Gaussian noise, which is inherent in many physical systems due to thermal fluctuations and other environmental factors.\par
	
	One of the most basic set of states is spin-coherent states, which are also called atomic coherent states, Bloch coherent states and angular momentum coherent states \cite{Aravind1999,Byrnes2015,Baguette2017}. These are a generalization of coherent states or Glauber states, which have a minimum uncertainty and can be produced classically by acting on the ground state \cite{Markham2003,TwarequeAli1995}. Recall that coherent states of a quantum mechanical harmonic oscillator are the eigenstates of its bosonic annihilation operator forming an overcomplete family. These states can be well understood in phase space within the framework of quantum distributions such as the Wigner function. This distribution was initially introduced by Wigner in 1932 as a consequence of his attempts to establish a quantum correction for thermodynamics \cite{Wigner1932}. The Wigner function has played a key role in the phase-space reformulation of quantum mechanics and has also potential applications mainly in statistical mechanics \cite{Lee1995, Rundle2021}, continuous variable quantum information \cite{Andersen2010, MistaJr2010}, quantum optics and scattering theory \cite{Iotti2017,Bastiaans1980}. Generally, it can be defined as a real-valued joint position-momentum quasiprobability, which may be non-positive for some particular states. Those states possessing negative Wigner functions are regarded as nonclassical states \cite{Cormick2006,Mandal2018}.
	
	The Wigner function and the Skew information are both involved in the study of quantum information, yet they have different applications and goals \cite{Frieden2006,Royer1989, Audenaert2008,Luo2004}. Despite this, Zhang and Luo have investigated in 2021 the intrinsic relations between these two concepts by exploiting symmetry and asymmetry of bosonic field states with respect to the displaced parity operator \cite{Zhang12021}. Here, parity operation simply means a reversal of spatial coordinates, which reads in the $n$-dimensional space as: 
	$\left(x_{1},x_{2},...,x_{n}\right)\longmapsto \left(-x_{1},-x_{2},...,-x_{n}\right)$. Driven by the intention to explore potential connections between the Wigner function and the Wigner-Yanase skew information, we have quantified in the present paper parity symmetry and asymmetry in terms of the Wigner function and Wigner-Yanase skew information, respectively. This work is structured as follows: In Section (\ref{sec2}), we review the concepts of Wigner function and Wigner-Yanase skew information and derive their relation by using the Wigner parity operator. Section (\ref{sec3}) is dedicated to an examination of parity symmetry and asymmetry in the superposition of the two spin-$1/2$ coherent states, as well as in a general spin-$j$ states.  Section (\ref{sec4}) is devoted to an investigation of parity symmetry and  asymmetry in the superposition of spin-$1/2$ state , as well as in a general spin-$j$ states under Gaussian noise channel. Finally, we conclude with a closing remarks.
	
	\section{Preliminaries} \label{sec2}
	Here, we review some basic concepts and relevant equations that we will use in subsequent sections. We will quantify and characterize nonclassicality using these two specific quantifiers: 
	\subsection{Wigner function}
	The Wigner function, a pivotal tool in quantum physics, detects nonclassical light behavior through its negativity. It has recently been employed in quantum information theory to identify quantum entanglement. It is also extensively used in various fields, such as time-frequency analysis and quantum optics. Its versatility and significance make it a ubiquitous tool in these areas. Initially designed for continuous variable quantum states, various methods have been devised to map discrete quantum systems onto a phase space framework within a discrete-dimensional Hilbert space. The Wigner function is defined as the Fourier transform of the non-diagonal elements of $\hat{\rho }$ in the basis of the eigenstates of the position and is given by
	\begin{equation}\label{nom2}
	W_{\hat{\rho}}(q,p)=\left(\frac{1}{2\pi\hbar }\right)^n\int_{-\infty}^{+\infty}du \hspace{0.1cm}  e^{-ip.u/ \hbar} \left\langle q+\frac{u}{2} \left\arrowvert \hat{\rho } \right \arrowvert q-\frac{u}{2} \right\rangle,
	\end{equation}
	where $q = [q_1, q_2,\dots, q_n]$ and $p = [p_1, p_2, \dots, p_n]$ are the classical phase-space position and momentum values (represented as $n$-dimensional vectors), $u = [u_1, u_2,\dots, u_n]$ and $\hbar$ is the constant of Planck. If the quantum state is a pure state, then the Wigner function becomes as follows
	\begin{equation}\label{nom3}
	W_{\hat{\rho}}(q,p)=\left(\frac{1}{2\pi\hbar }\right)^n\int_{-\infty}^{+\infty}du \hspace{0.1cm}  e^{ip.u/ \hbar} \psi^*\big(q+\frac{u}{2}\big)\psi\big(q-\frac{u}{2}\big).
	\end{equation}
	The Wigner function has several important properties that make it a valuable tool in quantum mechanics. It is a continuous function with real values (not necessarily positive), and satisfies the following conditions; It is bounded and does not have unlimited magnitude, as it is constrained by an upper limit; that is, 
		\begin{equation}
|W_{\hat{\rho}}(q,p)|\leq \left(\frac{1}{\pi\hbar }\right)^{n},
		\end{equation}
	it is also normalized, satisfying
		\begin{equation}
\int_{-\infty}^{+\infty}dq\int_{-\infty}^{+\infty} dp W_{\hat{\rho}}(q,p)= Tr[\hat{\rho}]=1,
		\end{equation}
and the trace product rule;
		\begin{equation}
Tr(\hat{\rho_1}\hat{\rho_2})=(2\pi\hbar)^n\int_{-\infty}^{+\infty}dq\int_{-\infty}^{+\infty} dp \  W_{\hat{\rho_1}}(q,p) \  W_{\hat{\rho_2}}(q,p).
		\end{equation}

The Wigner function can also be expressed in terms of  the displaced parity operator (also known as the Wigner kernel  or the Wigner operator) $\hat{\Delta}(q,p)$ according to \cite{He2022}: 
\begin{equation}\label{nomm3}
	W_{\hat{\rho}}(q,p)= Tr[\hat{\rho}\hat{\Delta}(q,p)],
	\end{equation}
such that the displaced parity operator $\hat{\Delta}(q,p)$, in terms of coordinate momentum variables, is given by
	\begin{equation}
	\hat{\Delta}(q,p)=\left(\frac{1}{2\pi\hbar }\right)^n\int_{-\infty}^{+\infty}du \hspace{0.1cm}  e^{-ip.u/ \hbar}  \arrowvert q-\frac{u}{2} \rangle \langle q+\frac{u}{2} \arrowvert.
	\end{equation}
	Furthermore for any complete parametrization $\Omega
	=\{q,p\}$ of the phase space, it fulfills the following constrained form of the Stratonovich-Weyl correspondence \cite{Tilma2016}:
	\begin{equation}
	\hat{\Delta}(\Omega)=\hat{\Delta}^+(\Omega),\hspace{1cm}\int_{\Omega}\hat{\Delta}(\Omega)d{\Omega}=1,\hspace{1cm}{\rm and}\hspace{1cm}\hat{\Delta}(\Omega')=U\hat{\Delta}(\Omega)U^{+},
	\end{equation}
	where $U$ is the unitary operator.
	\subsection{Wigner-Yanase skew information and its relation to the Wigner function}
	Quantum mechanics fundamentally limits the precision with which certain pairs of physical properties of a particle can be measured, unlike classical physics where any two observables can be measured with arbitrary accuracy. In quantum mechanics, the overall uncertainty resulting from the action of an observable $\hat{X}$ on a quantum state $\hat{\rho}$ is generally quantified using the variance \cite{Luo2005}, a measure defined by the following relation 
	\begin{equation}\label{variance}
	Var\big({\hat{\rho}},\hat{X}\big)=Tr\big({\hat{\rho}}\hat{X}^2\big)-\big(Tr\big({\hat{\rho}}\hat{X}\big)\big)^2.
	\end{equation}
	In particular, for the pure states $\hat{\rho}=|\psi\rangle \langle \psi|$, the variance reduces to 
	\begin{equation}\label{nomm2}
	Var(|\psi\rangle \langle \psi|,\hat{X})= \langle \psi| \hat{X}^2 |\psi\rangle - \langle \psi| \hat{X} |\psi\rangle^2 .
	\end{equation}   
	In the case of pure states, the variance is a well-defined measure of uncertainty. However, for mixed states, the variance is composed of two parts: a classical part due to the ignorance resulting from the mixed nature of the state, and a quantum part due to the non-commutativity between the state and the observable. To isolate the quantum part of the variance, Wigner and Yanase introduced the concept of skew information \cite{wigner1963}, defined by 
	\begin{equation}\label{nomm7}
	I(\hat{\rho},\hat{X}):= I_{\hat{\rho}}(\hat{X}):= \frac{1}{2}Tr\{(i[\sqrt{\hat{\rho}}, \hat{X}])^2\},
	\end{equation}  
	where $[.,.]$ is the commutator. Thus, we can define Wigner-Yanase skew information as a quantifier of the degree of non-commutativity between a quantum state $\hat{\rho}$ and an observable $\hat{X}$ that can be considered as a Hamiltonian or any other conserved quantity, and unlike variance, classical mixing has no effect on Wigner-Yanase skew information.\par
	
	In the particular case when $\hat{X}=\hat{\Delta}(q,p)$, the variance (\ref{variance}) can be expressed using the Wigner function as follows 
	
\begin{equation}\label{nom2}
	Var({\hat{\rho}},\hat{\Delta}(q,p))= 1-W_{\hat{\rho}}(q,p)^2,
	\end{equation}
	since $\hat{\Delta}(q,p)^2=1$ and $W_{\hat{\rho}}(q,p)= Tr[\hat{\rho}\hat{\Delta}(q,p)]$.\par
	
	On the other hand, in the Wigner-Yanase skew information $I(\hat{\rho},\hat{X})$ we have the freedom to choose a reference observable $\hat{X}$, so we can consider $\hat{X}$ as  the displaced parity operator $\hat{\Delta}(q,p)$, and then   
	\begin{equation}\label{nomm4}
	I(\hat{\rho},\hat{\Delta}(q,p))= -\frac{1}{2}Tr[\sqrt{\hat{\rho}},\hat{\Delta}(q,p)]^{2},
	\end{equation}
	where the Wigner-Yanase skew information is always dominated by the variance. Consequently, We have the following relationship between Wigner function and Wigner-Yanase skew information \cite{Zhang12021}
\begin{equation} 
 I(\hat{\rho},\hat{\Delta}(q,p))+  {W_{\hat{\rho}}}^2(q,p) \leq 1, \label{nomm18}
 \end{equation}
 
where it underscores the interplay between symmetry and asymmetry in quantum states, suggesting that the properties of the Wigner function can offer insights into the information-theoretic aspects of quantum states as captured by the Wigner-Yanase skew information. This inequality, as expressed in equation (\ref{nomm18}), becomes an equality for pure states, representing the conservation relation between the Wigner function and the Wigner-Yanase skew information, i.e.,
\begin{equation}\label{nomm12} I(\rho,\hat{\Delta}(q,p))+  {W_{\hat{\rho}}}^2(q,p)=1. 
 \end{equation}
 
As a physical interpretation of the relations (\ref{nomm18}) and (\ref{nomm12}), the magnitude of the mean value and the quantum uncertainty of the kernel operators in each state satisfy the constraint relations. This interpretation arises from the fact that the Wigner function represents the mean value of these kernel operators when expressed in terms of them, and from the understanding that the Wigner-Yanase skew information, which incorporates kernel operators, provides a physical interpretation of quantum uncertainty. This relationship exemplifies the concept of symmetry-asymmetry complementarity: the square of the Wigner function illustrates the symmetry of the state with respect to the kernel operators, while the Wigner-Yanase skew information quantifies the asymmetry of the state in relation to those operators. This quantification provides a clearer understanding of how quantum states deviate from symmetry, which is crucial for characterizing nonclassical states in quantum optics. The findings of Ref.\cite{Zhang12021} suggest that the degree of parity asymmetry is not merely a mathematical construct but has tangible physical consequences. For example, it can be correlated with the quantumness of a state, a crucial property for applications in quantum information processing. This implies that a deeper understanding of parity asymmetry could lead to valuable insights into the nonclassical characteristics of quantum states, which are essential for advancing quantum technologies.

\section{The parity symmetry of spin coherent states}\label{sec3}
 Coherent states, often referred to as Glauber states, are quantum mechanical states of a harmonic oscillator that closely resemble classical behavior \cite{Gazeau2009,Glauber1963}. They are fundamental to quantum optics, especially laser physics, and have been extensively studied by Roy J. Glauber \cite{GLAUBER1963}. Mathematically, coherent states are defined as the eigenstates of the annihilation operator, denoted by $a$. For a coherent state $|\alpha \rangle$, we have the following relationship
	\begin{equation}
	\begin{array}{ll}
	a|\alpha \rangle= \alpha |\alpha \rangle, \hspace{2cm} \alpha = |\alpha| e^{i\phi} \in \mathbb{C}. \end{array}
	\end{equation}
	This coherent states can be expressed in Fock space as
	\begin{equation}
	\begin{array}{ll}
	|\alpha \rangle =\exp{[-\frac{1}{2}|a|^2]} \sum_{n=0}^{\infty} \frac{a^n}{\sqrt{n !}}|n\rangle=\exp{[-\frac{1}{2}|a|^2]} \sum_{n=0}^{\infty} \frac{\left(\alpha a^+\right)^n}{n !}|0\rangle.
	\end{array}
	\end{equation}
	
	Here we aim to elucidate the behavior of parity symmetry and asymmetry of standard spin coherent states, also known as atomic-named coherent states. These states play a fundamental role in exploiting and characterizing the quantum properties of spin systems. They are commonly referred to as Bloch states. These characteristic states represent a coherent superposition of different spin states in Hilbert space \cite{Monroe1996, Zhang1990} and provide an elegant and powerful mathematical representation of the direction of rotation of a particle, allowing a deeper understanding of the quantum complexities underlying spin motion and evolution. A distinctive feature of correlated spin states is their ability to reduce the uncertainty associated with measuring the direction of rotation: Their coherent organization produces distribution properties that uniformly encompass the entire Bloch sphere, giving it enhanced sensitivity to the smallest deviations in rotation angles. These spin coherent states are  defined as follows
	\begin{equation}
	\begin{array}{ll}
	\big|\theta, \varphi, j \big\rangle :=R(\theta,\varphi) \big|j,-j \big\rangle,
	\end{array}
	\end{equation}
	where $j$ is the spin value and $ R(\theta,\varphi):=\exp \left[\frac{\theta}{2}\left(J_{+} e^{-i \varphi}-J_{-}e^{i \varphi}\right)\right]$ is the rotation operator defined by generators $J_{\pm}$ (raising and lowering generators) and $J_z$ (the weight operator or Cartan) which generates $SU(2)$ algebra, obeying the following commutation relations $[J_{+}, J_{-}] = 2J_{z}$ and $[J_{z}, J_{\pm}] = \pm J_{\pm}$. Moreover;
	\begin{equation}
	J_{+}|j, m \big\rangle =\sqrt{(j-m)(j+m+1)}|j, m+1 \big\rangle, \hspace{0.5cm} m\neq j,
	\end{equation}
and
\begin{equation}
J_{-}\big|j, m \big\rangle =\sqrt{\big(j-m+1\big)\big(j+m\big)} \big|j, m-1 \big\rangle, \hspace{0.5cm}   m\neq-j.
\end{equation}
	Furthermore, the spin coherent state can be expressed in the Dicke basis as 
	\begin{align}
	\big|\theta, \varphi, j \big\rangle= \sum_{m=-j}^j\left(\begin{array}{c}
	2 j \\
	j+m
	\end{array}\right)^{\frac{1}{2}} \bigg(\cos\bigg(\frac{\theta}{2}\bigg)\bigg)^{j-m} \bigg(e^{-i\varphi}\sin\bigg(\frac{\theta}{2}\bigg)\bigg)^{j+m} \big|j, m\big\rangle, 
	\end{align}
	where the Dicke states $\big|j, m \big\rangle$ correspond to the two-mode Fock states
	\begin{equation}
\big|j, m \big\rangle=\big|j+m \big\rangle \otimes \big|j-m \big\rangle,\hspace{0.5cm} m=-j,-j+1,...,j-1,j.
	\end{equation}
In this context, $j$ represents the total spin quantum number, while m denotes its projection along the z-axis. These states are associated with the common eigenstates of the commuting operators $J^2$ and $J_z$, which satisfy the following equation:
\begin{equation}
J_z\big|j, m \big\rangle= m\big|j, m \big\rangle,\hspace{0.5cm} J^2\big|j, m \big\rangle=j(j+1)\big|j, m \big\rangle.
\end{equation}
This construction is based on the Schwinger representation of the $SU(2)$ Lie algebra, whereby the angular momentum operators are expressed in terms of bosonic creation and annihilation operators as follows \cite{Zhang2021}
\begin{equation}
    J_x= \frac{1}{2}(a_1^{\dagger}\otimes a_2 +a_1\otimes a_2^{\dagger}), \hspace{0.5cm}   J_y= \frac{i}{2}(a_1\otimes a_2^{\dagger} -a_1^{\dagger}\otimes a_2), \hspace{0.5cm} J_z= \frac{1}{2}(a_1^{\dagger}a_1 \otimes \mathbf{1}-\mathbf{1}\otimes a_2^{\dagger}a_2).
\end{equation}
These operators satisfy the commutation relations of $SU(2)$:
\begin{equation}
    [J_x,J_y]=i J_z, \hspace{0.5cm} [J_y,J_z]=i J_x, \hspace{0.5cm} [J_z,J_x]=i J_y.
\end{equation}
The ladder operators $J_+$ and $J_-$
 are defined as:
 \begin{equation}
     J_+=J_x+iJ_y, \hspace{0.5cm} J_-=J_x -iJ_y,
 \end{equation}
  these operators permit transitions between states $\big|j, m \big\rangle$ by raising or lowering the $m$-value, respectively.
\subsection{Parity symmetry and asymmetry of spin-$1/2$ cat states}
In the Fock state basis, when $j = 1/2$, it signifies the presence of just a single photon in the coherent state. In this scenario, the coherent state can be expressed as
	\begin{equation}
	\big|\theta, \varphi, \frac{1}{2}\big\rangle= \cos\big(\frac{\theta}{2}\big)\big|\frac{1}{2}, -\frac{1}{2}\big\rangle + e^{-i\varphi}\sin\big(\frac{\theta}{2}\big)\big|\frac{1}{2}, \frac{1}{2}\big\rangle.
	\end{equation}
	Throughout this framework, we represent the general form of the superposition involving two spin coherent states as: 
	\begin{align}
	\big|Cat, \frac{1}{2}\big\rangle &= \mathcal{N} \bigg(\big|\theta_1, \varphi_1, \frac{1}{2}\big\rangle + \big|\theta_2, \varphi_2,\frac{1}{2}\big\rangle\bigg)\notag
	\\&=\mathcal{N} \bigg(\big(\cos\big(\frac{\theta_1}{2}\big)+\cos(\frac{\theta_2}{2})\big)\big|\frac{1}{2}, -\frac{1}{2}\big\rangle + \big(e^{-i\varphi_1}\sin\big(\frac{\theta_1}{2}\big)+e^{-i\varphi_2}\sin\big(\frac{\theta_2}{2}\big)\big)\big|\frac{1}{2}, \frac{1}{2}\big\rangle \bigg),
	\end{align}
	where $\mathcal{N}$ represents the normalization factor of the spin cat states, and it can be determined as
	\begin{equation}
	\mathcal{N}={\bigg(2+2\cos(\frac{\theta_1}{2})\cos(\frac{\theta_2}{2})+2\cos(\varphi_1-\varphi_2)\sin(\frac{\theta_1}{2})\sin(\frac{\theta_2}{2})\bigg)}^{-\frac{1}{2}}
	\end{equation}
 
Now, we are interested in determining the symmetry or asymmetry of a state with respect to the displaced parity operator. So, because the square of the Wigner function is interpreted as a quantifier of the symmetry state with respect to the displaced parity operator, thus the displaced parity operator of $\big|Cat, \frac{1}{2}\big\rangle$ can be expressed using the standard displacement operator $\hat{D}(\alpha):=\exp(\alpha a_1^{\dagger} -\alpha a_1)$ and $\hat{D}(\beta):=\exp(\beta a_2^{\dagger} -\beta a_2)$, and the parity operator, given by $\hat{\Pi}_{1}:=(-1)^{a_1^{\dagger}a_1}$ and $\hat{\Pi}_{2}:=(-1)^{a_2^{\dagger}a_2}$ as shown in the following equality
	\begin{align}\label{nomm21}
\hat{\Delta}\big(q_1,q_2,p_1,q_2\big):=\hat{\Delta}\big(\alpha,\beta\big)&=\left(\frac{1}{\pi\hbar }\right)^2\int_{-\infty}^{+\infty}\Big[ e^{-2ip_1.u_1/ 2\hbar}  e^{-2ip_2.u_2/ 2\hbar}  D(\alpha) (-1)^{a_1^{\dagger}a_1} \arrowvert \frac{u_1}{2} \rangle \langle \frac{u_1}{2} \arrowvert D(\alpha)^{\dagger} D(\beta) \\& \nonumber \times (-1)^{a_2^{\dagger}a_2} \arrowvert \frac{u_2}{2} \rangle \langle \frac{u_2}{2} \arrowvert D(\beta)^{\dagger}\Big] d\big(\frac{u_1}{2}\big) d\big(\frac{u_2}{2}\big),
	\end{align}
where $$\alpha=\frac{q_1 + i p_1 }{\sqrt{2}} \in \mathbb{C}, \hspace{0.5cm}   \hspace{0.5cm} 
\beta=\frac{q_2 + i p_2 }{\sqrt{2}}\in \mathbb{C}, \hspace{0.5cm} p_1,q_1,p_2,q_2\in \mathbb{R}$$ represent the complete parameterization of the phase space. For any complete parameterization $\alpha$ or $\beta$ of the phase space, such that $\hat{D}$ and $\hat{\Pi}$ are defined in terms of coherent states $\hat{D}(\alpha) \arrowvert 0\rangle=\arrowvert \alpha\rangle$ and $\hat{D}(\beta) \arrowvert 0\rangle=\arrowvert \beta\rangle$, wile $\hat{\Pi}_{1}\arrowvert \alpha\rangle=\arrowvert -\alpha\rangle$ and $\hat{\Pi}_{2}\arrowvert \beta\rangle=\arrowvert -\beta\rangle$. In this situation, the displacement operator $\hat{D}$ is often parametrized in terms of position and momentum coordinates or eigenvalues of the annihilation operators.\par 

Then, by  $(\ref{nomm21})$, the displaced parity operator in  the phase space of quantum optics of two orthogonal harmonic oscillators can be properly described as the product of the corresponding displaced parity operators:
	\begin{equation}\label{nom22}
	\hat{\Delta}(\alpha,\beta)= \hat{\Delta}(\alpha)\hat{\Delta}(\beta),
	\end{equation}
where $\hat{\Delta}(\alpha)$ and $\hat{\Delta}(\beta)$ are the displaced parity operators of single-mode harmonic oscillator in the $x$ and $y$ directions,  respectively.\par
In the following we will work  on the two mode. So we can use the expression  (\ref{nomm3}) to achieve the pparity-symmetric behavior of the spin-1/2 cat state. Therefore, we obtain the subsequent function
	\begin{align}\label{Wigner function1}
	 W_{\mid Cat, \frac{1}{2}\rangle \langle\frac{1}{2}, Cat\mid}\big(q_1,q_2,p_1,p_2\big) &:=W_{\mid Cat, \frac{1}{2}\rangle \langle\frac{1}{2}, Cat\mid}\big(\alpha,\beta\big)\\& \nonumber  = \bigg(\frac{\mathcal{N}}{\pi}\bigg)^2 \bigg[\bigg(\cos\bigg(\frac{\theta_1}{2}\bigg)+ \cos\bigg(\frac{\theta_2}{2}\bigg)\bigg)^2 e^{-2|\alpha|^2} e^{-2 \mid 1-\beta \mid^2} +\bigg(\cos\bigg(\frac{\theta_1}{2}\bigg)+ \cos\bigg(\frac{\theta_2}{2}\bigg)\bigg)
	\\& \nonumber\times \bigg(e^{-i \varphi_1} \sin\bigg(\frac{\theta_1}{2}\bigg)+e^{-i \varphi_2} \sin(\frac{\theta_2}{2})\bigg) e^{-\frac{1}{2} \mid 2 \alpha-1\mid^2- \alpha+\alpha^\ast} e^{-\frac{1}{2}\mid 1- 2\beta\mid^2 + \beta - \beta^\ast} \\& \nonumber + \bigg(\cos\bigg(\frac{\theta_1}{2}\bigg)+ \cos\bigg(\frac{\theta_2}{2}\bigg)\bigg)
	\bigg(e^{i \varphi_1} \sin(\frac{\theta_1}{2})+e^{i \varphi_2}\sin\bigg(\frac{\theta_2}{2}\bigg)\bigg)  e^{-\frac{1}{2} \mid 2 \beta-1\mid^2- \beta+\beta^\ast}  e^{-\frac{1}{2}\mid 1- 2\alpha\mid^2 + \alpha - \alpha^\ast}
	\\& \nonumber+  \bigg(\sin^2\bigg(\frac{\theta_1}{2}\bigg)+2 \sin\bigg(\frac{\theta_1}{2}\bigg) \sin\bigg(\frac{\theta_2}{2}\bigg)\cos\bigg(\varphi_1-\varphi_2\bigg)+\sin^2\bigg(\frac{\theta_2}{2}\bigg)\bigg) e^{-2\mid 1- \alpha\mid^2}  e^{-2\mid \beta \mid^2}\bigg],
	\end{align}

 \begin{figure}[hbtp]
  			{{\begin{minipage}[b]{.2\linewidth}
  						\centering
  						\includegraphics[scale=0.30]{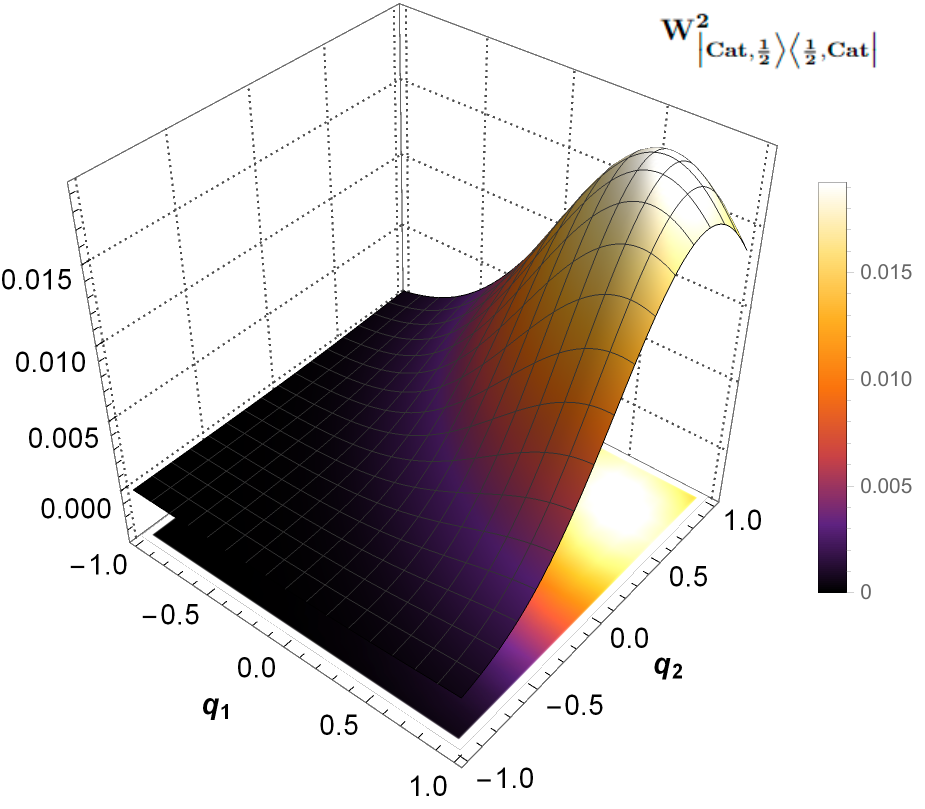} \vfill $\left(a\right)$
  					\end{minipage} \hfill
  					\begin{minipage}[b]{.2\linewidth}
  						\centering
  						\includegraphics[scale=0.30]{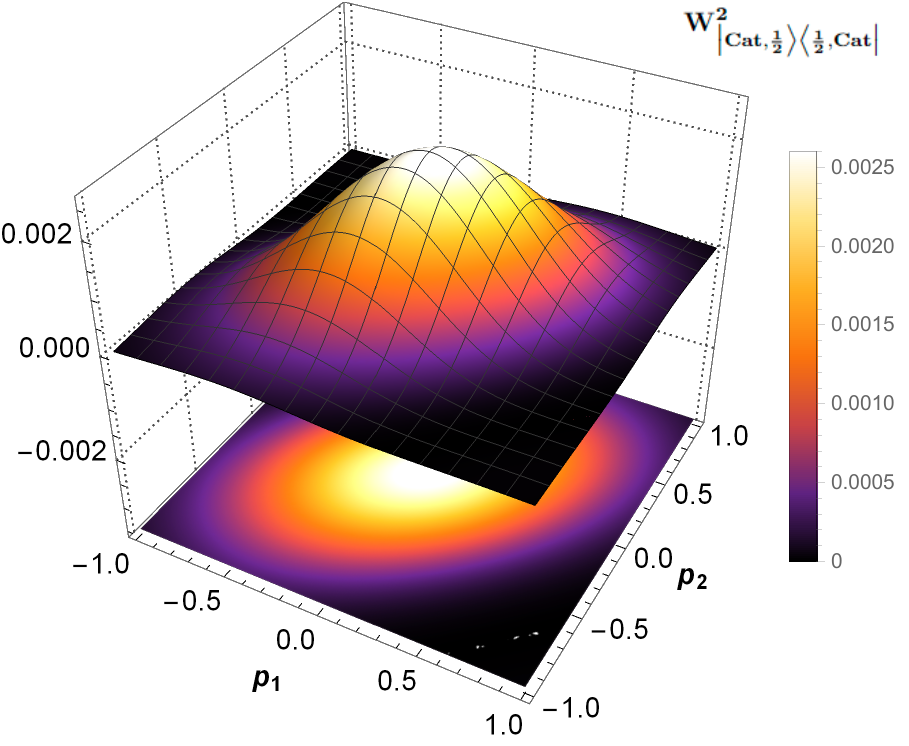} \vfill  $\left(b\right)$
 					\end{minipage} \hfill
  					\begin{minipage}[b]{.2\linewidth}
  						\centering
  						\includegraphics[scale=0.30]{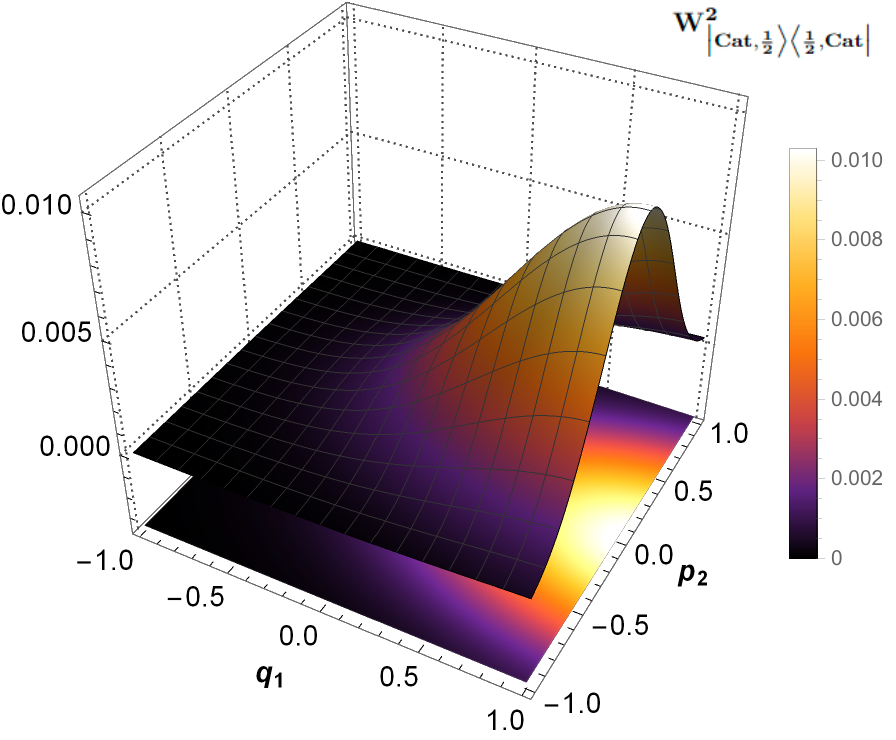} \vfill  $\left(c\right)$
  			\end{minipage} \hfill
     \begin{minipage}[b]{.2\linewidth}
  						\centering
  						\includegraphics[scale=0.30]{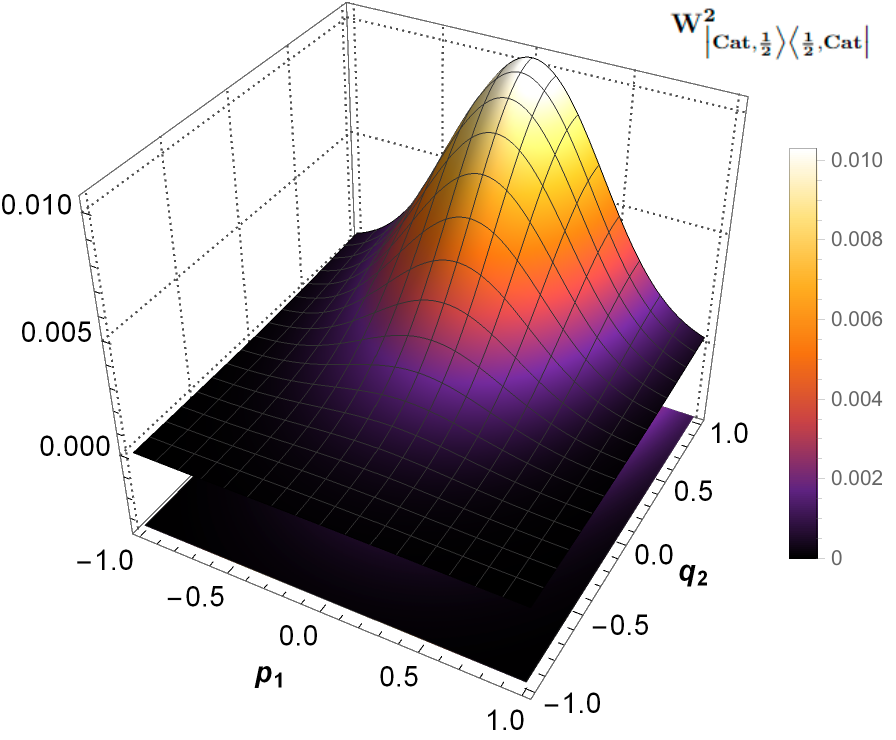} \vfill $\left(d\right)$
  				\end{minipage}}}
      {{\begin{minipage}[b]{.2\linewidth}
  						\centering
  						\includegraphics[scale=0.30]{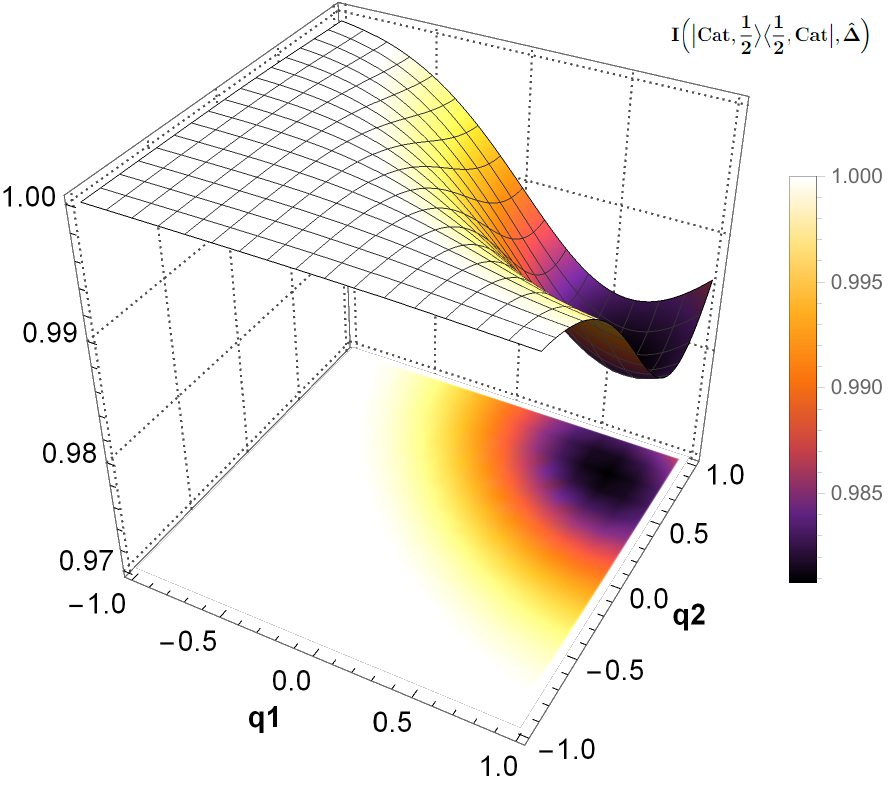} \vfill $\left(e\right)$
  					\end{minipage} \hfill
  					\begin{minipage}[b]{.2\linewidth}
  						\centering
  						\includegraphics[scale=0.30]{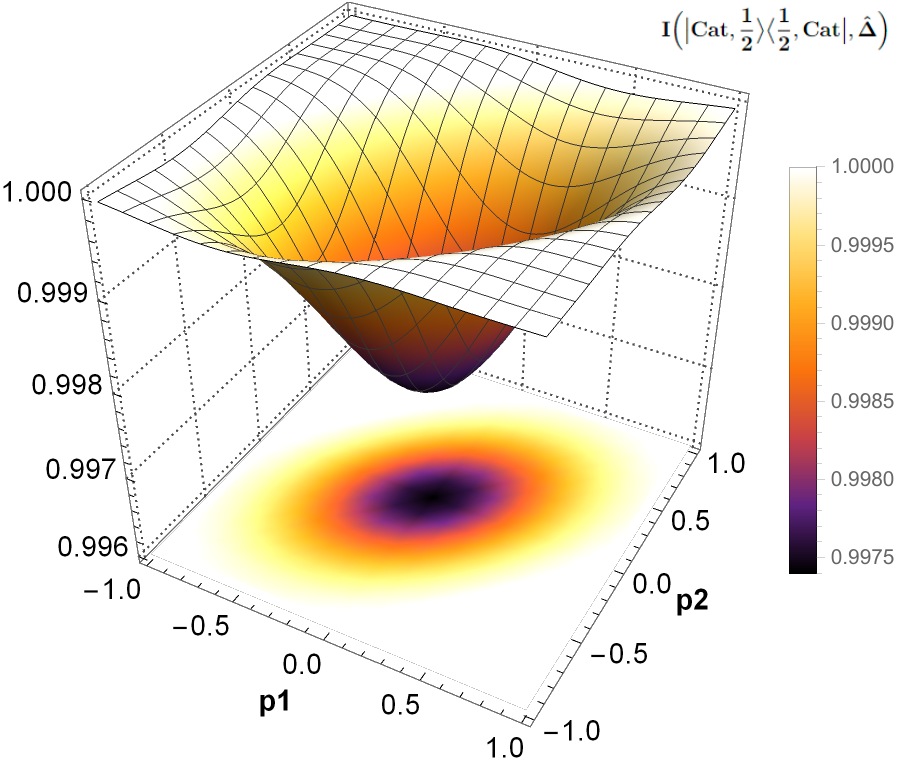} \vfill  $\left(f\right)$
 					\end{minipage} \hfill
  					\begin{minipage}[b]{.2\linewidth}
  						\centering
  						\includegraphics[scale=0.30]{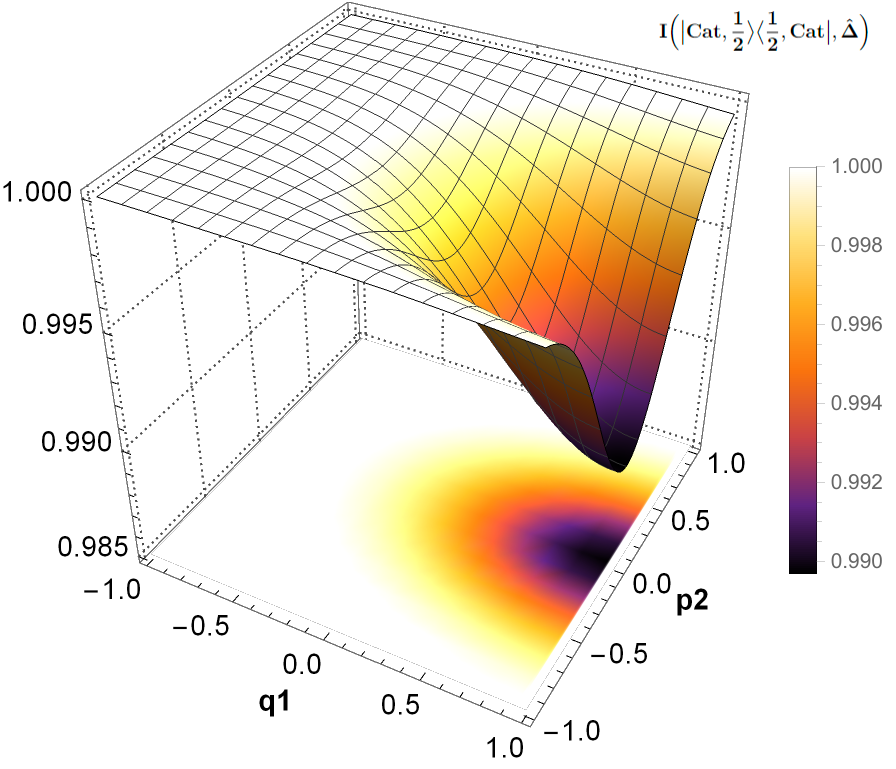} \vfill  $\left(g\right)$
  			\end{minipage} \hfill
     \begin{minipage}[b]{.2\linewidth}
  						\centering
  						\includegraphics[scale=0.30]{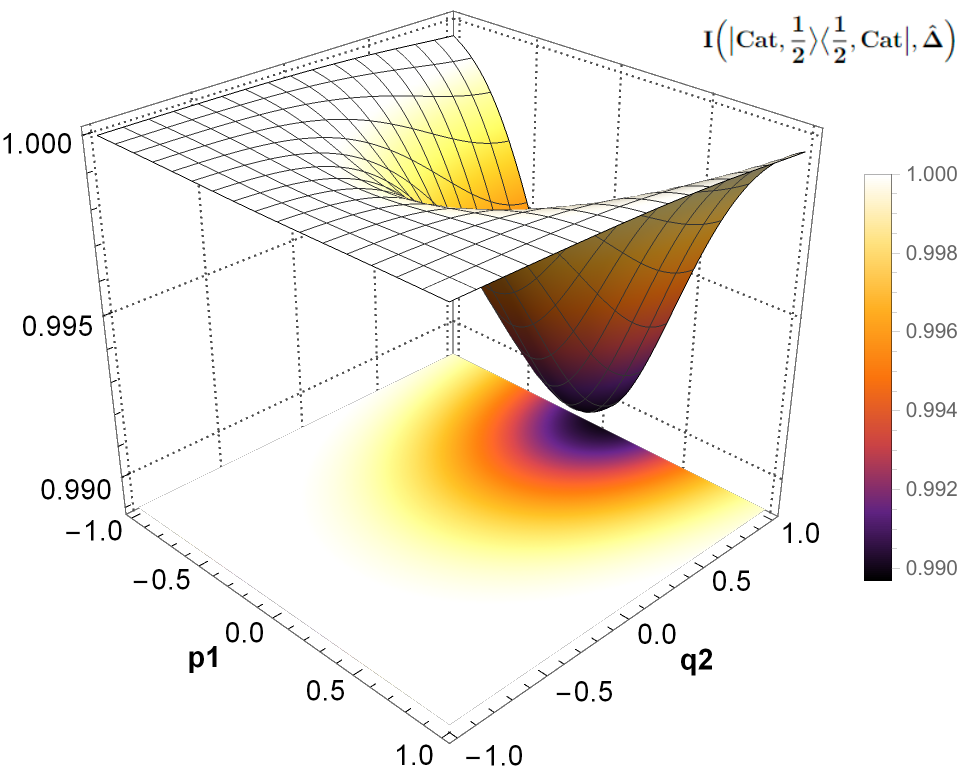} \vfill $\left(h\right)$
  				\end{minipage}}}
  \caption{Square of the Wigner function (\ref{Wigner function1}) and Wigner-Yanase skew information (\ref{nomm12}) of spin-$1/2$ cat state in phase space: \textbf{(a)} Square of the Wigner function and \textbf{(e)} Wigner-Yanase skew information for $p_1=p_2=0$, \textbf{(b)}  Square of the Wigner function and \textbf{(f)} Wigner-Yanase skew information for $ q_1=q_2=0$, \textbf{(c)}  Square of the Wigner function and \textbf{(g)} Wigner-Yanase skew information for $p_1=q_2=0 $,  \textbf{(d)}  Square of the Wigner function and \textbf{(h)} Wigner-Yanase skew information for $ q_1=p_2=0$; obtained for the fixed values $\theta_1=\pi$, $\theta_2=0$, $\phi_1=0$ and $\phi_2= 2\pi$.}
	    \label{fig1}
 	 	\end{figure}
    
We study the behavior of parity symmetry and asymmetry in the spin-$1/2$ cat state by using the conservation relation for the Wigner function and the Wigner-Yanase skew information concerning pure states in equation (\ref{nomm18}). Fig.\ref{fig1} show 3D the dependence of the square of the Wigner function and the Wigner-Yanase skew information on both the real and imaginary parts of $\alpha$ and $\beta$; with $p_1=p_2=0$ (Fig.$\ref{fig1}$ $\left(a\right)$ and $\left(e\right)$), $q_1=q_2=0$ (Fig.$\ref{fig1}$ $\left(b\right)$ and $\left(f\right)$), $p_1=q_2=0$ (Fig.$\ref{fig1}$ $\left(c\right)$ and $\left(g\right)$) and $q_1=p_2=0$ (Fig.$\ref{fig1}$ $\left(d\right)$ and $\left(h\right)$).\par
The results show that the symmetry and asymmetry of $|Cat, 1/2\rangle$ with respect to the displaced parity operator tend to reach the maximum value $1$ in certain regions of the real and imaginary parts of $\alpha$ and $\beta$. This confirms the violation of parity symmetry when the Wigner-Yanase skew information  reaches its maximum value and the preservation of parity symmetry when the square of the Wigner function equals $1$.\par
According to Fig.\ref{fig1} $\left(a\right)$ and $\left(e\right)$, the parity asymmetry is minimal for $q_2 > 0$ and $q_1 > 0$. In regions far from $q_2 > 0$ and $q_1 > 0$, the asymmetry of $|Cat, 1/2\rangle$ with respect to $\hat{\Delta}(\alpha,\beta)$ tends increasingly toward the maximum value  $1$, which is the opposite for the parity symmetry. For example, by taking $q_1 = -1$ and $q_2 = -1/2$, the parity asymmetry reaches its maximum value, whereas if $q_1 = 1$ and $q_2 = 1/2$, the parity asymmetry reaches its minimum value. In Fig.\ref{fig1} $\left(b\right)$ and  $\left(f\right)$, the parity asymmetry tends to its maximum value when $-1 \leq p_2 \leq 0$ and $0 \leq p_1 \leq 1$,  when  $p_2$ and  $p_1$ are near $1$ or $-1$ and also when $-1 \leq p_1 \leq  -1/2$ and $1/2 \leq  p_2 \leq 1$. Fig.\ref{fig1} $\left(c\right)$ and $\left(g\right)$ show that the optimal value of the parity asymmetry yields when $p_2 = -1$ and  $p_2 = 1$, or $-1 \leq q_1 \leq  -1/2$. Similarly, in the regions where $p_1 = -1$ and  $p_1 = 1$,  or $-1 \leq q_2 \leq -1/2$, as depicted in Fig.\ref{fig1} $\left(d\right)$ and $\left(h\right)$.\par
Furthermore, it is known that the skew information is a version of the Fisher information $F(\hat{\rho} ,\hat{X})$, which represents a fundamental limit on the accuracy of estimating an unknown parameter generated by a unitary dynamics $U= exp(-i\hat{X}\theta)$, which plays a primary role in quantum metrology \cite{Giovannetti2011,Saidi2024, Cappellaro2005,AbouelkhirH2023}. The ultimate limit on parameter estimation is given by the quantum-unbiased Cramér-Rao bound (CRB) ${\Delta\theta}^2F(\hat{\rho} ,\hat{X})\geq 1$, where $\Delta\theta$ characterises the accuracy of estimation by any possible measurement performed on the quantum state $U\hat{\rho}U^{\dagger}$. This makes it possible to consider the skew information or the square of the Wigner function as a fundamental bound on the precision of the estimation of an unknown parameter. To clarify, this amounts to replacing the Fisher information by the skew information or the square of the Wigner function in the Cramér-Rao bound. In this sense we can say that the Cramér-Rao bound has minimal values in the regions where the skew information is maximal, and minimal values for the square of the Wigner function, since it is the dual of the skew information, in particular for $(\theta_1,\theta_2)=(\pi,0)$ and $(\phi_1,\phi_2)=(0, 2\pi)$. This corresponds to the selected positions and momentum (the other quadrature variables are set to zero).  
\subsection{Parity symmetry of spin-$j$ cat states} 
Since the symmetry and the asymmetry are dual (as we saw in the previous subsection), it is sufficient to restrict our study in this subsection to the behavior of the parity symmetry of the spin coherent states. Then, we focus on the spin value $j$ and its impact on parity symmetry violation. For a more intuitive understanding, we consider the square of the Wigner function of the general form of the superposition of two spin-coherent states that can be written as follows 
	\begin{align}
	|Cat, j\rangle &= \mathcal{N}_t \bigg(|\theta_1, \varphi_1, j\rangle + |\theta_2, \varphi_2, j\rangle\bigg)
	\\& \nonumber= \mathcal{N}_t\sum_{m=-j}^j\left(\begin{array}{c}
	2 j \\
	j+m
	\end{array}\right)^{\frac{1}{2}} \left[\cos\bigg(\frac{\theta_1}{2}\bigg)^{j-m}\bigg(e^{-i\varphi_1}\sin\bigg(\frac{\theta_1}{2}\bigg)\bigg)^{j+m} + \cos\bigg(\frac{\theta_2}{2}\bigg)^{j-m}\bigg(e^{-i\varphi_2}\sin\bigg(\frac{\theta_2}{2}\bigg)\bigg)^{j+m}\right]|j, m\rangle,
	\end{align}
	where $\mathcal{N}_t$ is the normalization factor defined by
 \begin{align}
	\mathcal{N}_t &= \bigg(2+\bigg(\cos\bigg(\frac{\theta_1}{2}\bigg)\cos\bigg(\frac{\theta_2}{2}\bigg)+ \exp{[-i(\varphi_2-\varphi_1)]}\sin\bigg(\frac{\theta_1}{2}\bigg)\sin\bigg(\frac{\theta_2}{2}\bigg)\bigg)^{2j}+\bigg(\cos\bigg(\frac{\theta_1}{2}\bigg)\cos\bigg(\frac{\theta_2}{2}\bigg) \\& \nonumber \hspace{1cm}+ \exp{[i(\varphi_2-\varphi_1)]}\sin\bigg(\frac{\theta_1}{2})\sin\bigg(\frac{\theta_2}{2}\bigg)\bigg)^{2j}\bigg)^{-\frac{1}{2}}.
\end{align}
 
\begin{figure}[hbtp]
 			{{\begin{minipage}[b]{.2\linewidth}
 						\centering
 						\includegraphics[scale=0.30]{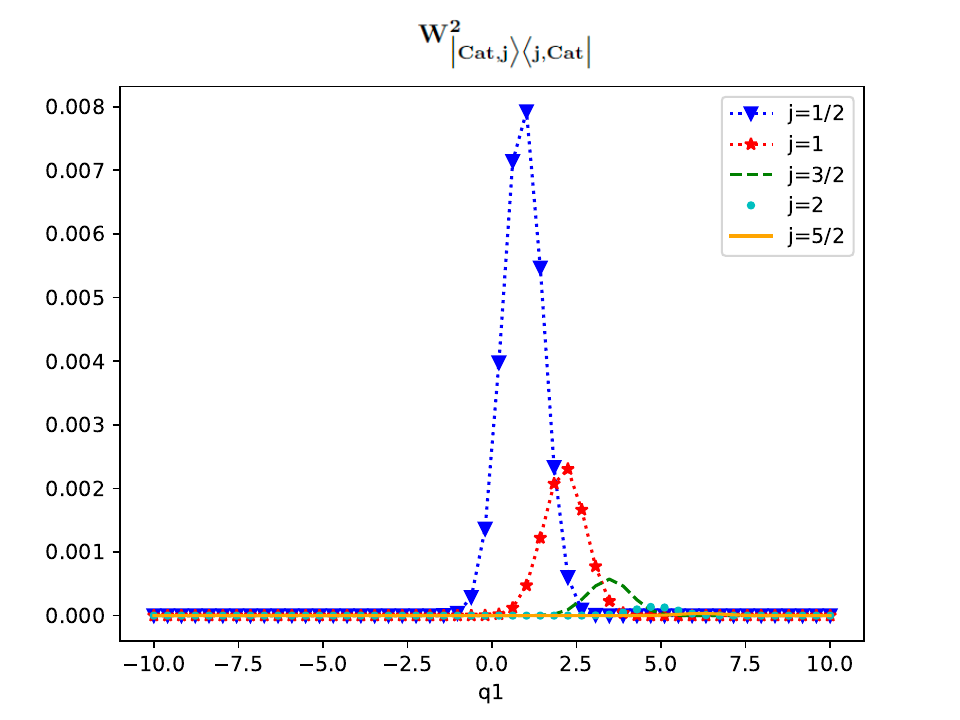} \vfill $\left(a\right)$
 					\end{minipage} \hfill
 					\begin{minipage}[b]{.2\linewidth}
 						\centering
 						\includegraphics[scale=0.30]{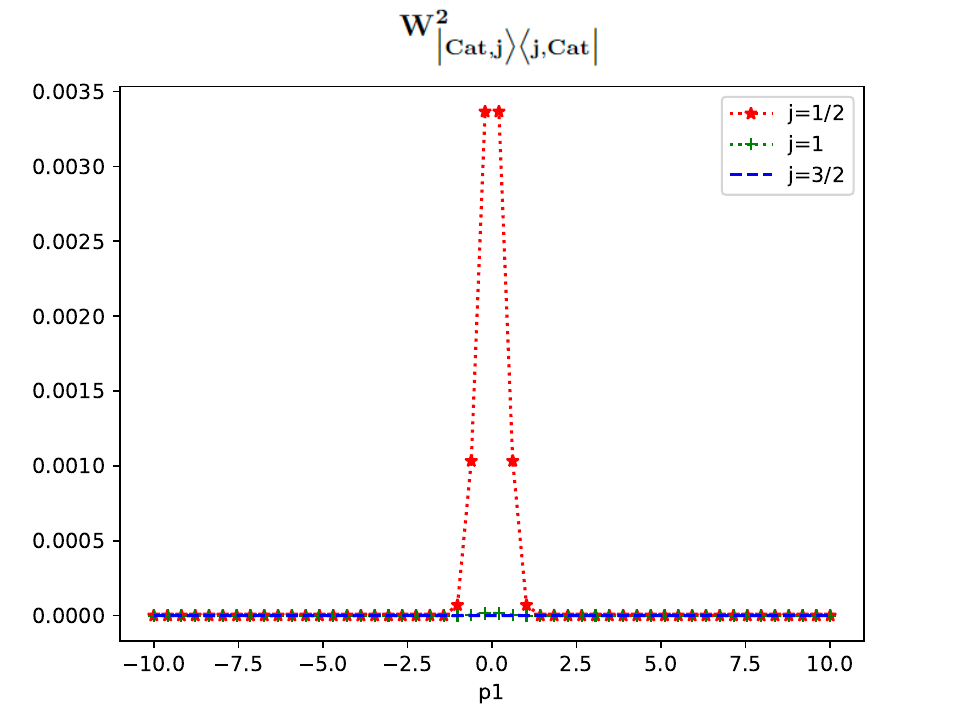} \vfill  $\left(b\right)$
 					\end{minipage} \hfill
 					\begin{minipage}[b]{.2\linewidth}
 						\centering
 						\includegraphics[scale=0.30]{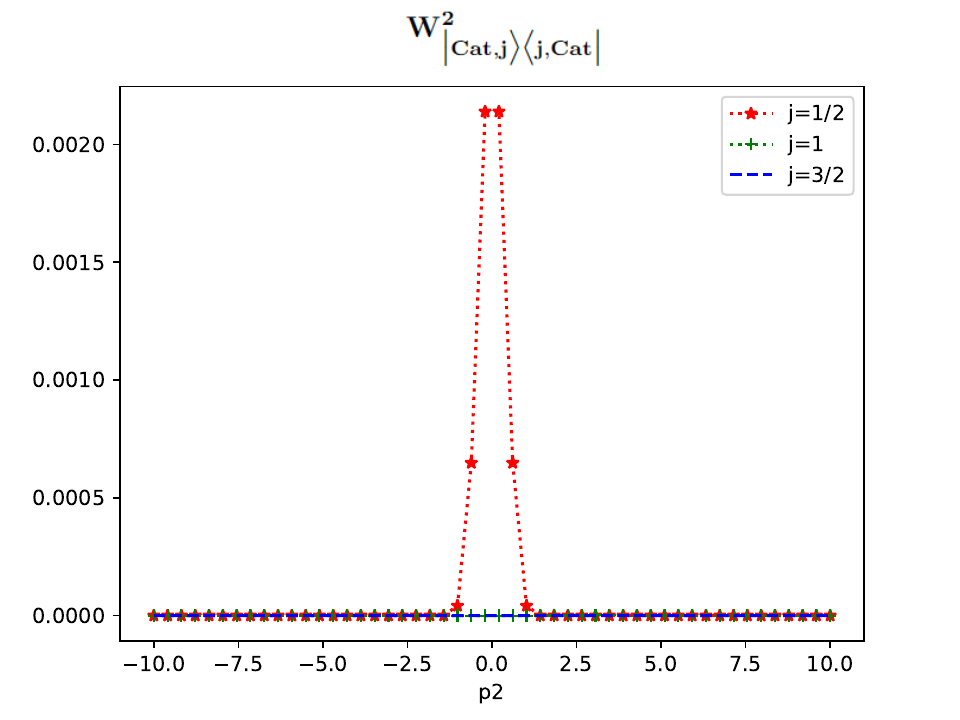} \vfill  $\left(c\right)$
 			\end{minipage} \hfill
    \begin{minipage}[b]{.2\linewidth}
 						\centering
 						\includegraphics[scale=0.30]{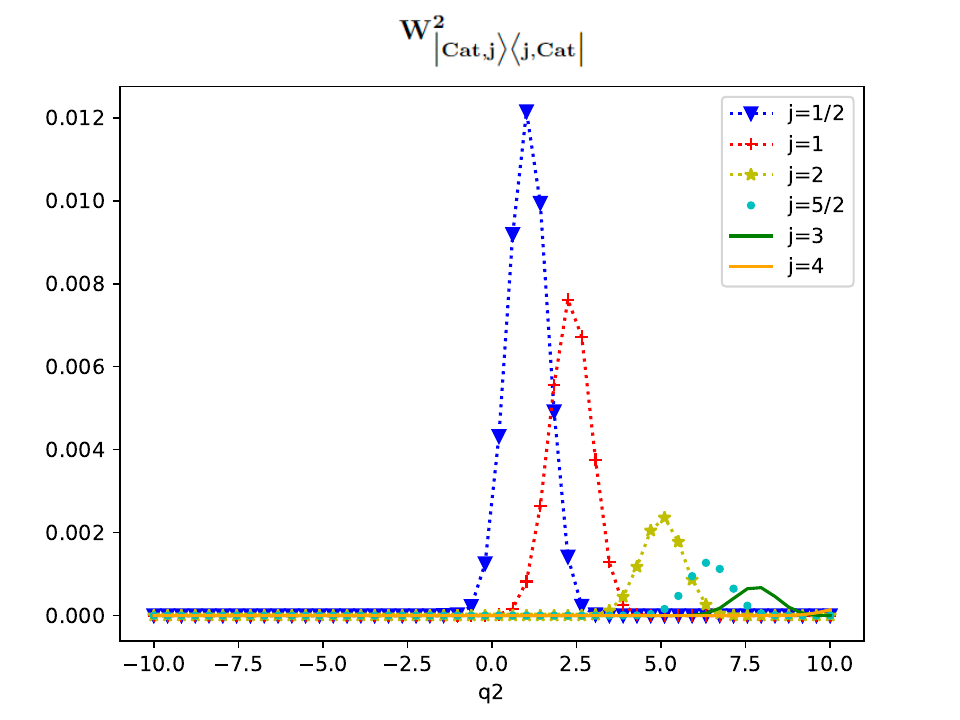} \vfill $\left(d\right)$
 					\end{minipage}}}
 \caption{The square of the Wigner function (\ref{nom28}) of spin-$j$ cat state in terms of $q_1$, $p_1$, $q_2$ and  $p_2$ for  different values of spin $j$.}
		\label{fig2}
 		\end{figure}

Consequently, we can find the following function, for $\theta_1$, $\theta_2$ $\in$ $]0,\pi[$, for any spin value $j$:
	\begin{align}\label{nom28}  
	W_{\mid Cat,j\rangle \langle j, Cat\mid}\big(q_1,q_2,p_1,p_2\big) &:= W_{\mid Cat, j\rangle \langle j, Cat\mid}\big(\alpha,\beta\big)\\& \nonumber  = \left(\frac{\mathcal{N}_t}{\pi}\right)^2\sum_{m,n=-j}^j\left(\begin{array}{c}
	2 j \\
	j+m
	\end{array}\right)^{\frac{1}{2}} \left(\begin{array}{c}
	2 j \\
	j+n 
	\end{array}\right)^{\frac{1}{2}}
	\bigg(\cos\bigg(\frac{\theta_1}{2}\bigg)^{j-m}\bigg(e^{i\varphi_1}\sin\bigg(\frac{\theta_1}{2}\bigg)\bigg)^{j+m}+\cos\bigg(\frac{\theta_2}{2}\bigg)^{j-m}\\& \nonumber\times\bigg(e^{i\varphi_2}\sin\bigg(\frac{\theta_2}{2}\bigg)\bigg)^{j+m}\bigg)\bigg(\cos\bigg(\frac{\theta_1}{2}\bigg)^{j-n}\bigg(e^{-i\varphi_1}\sin\bigg(\frac{\theta_1}{2}\bigg)\bigg)^{j+n}+\cos\bigg(\frac{\theta_2}{2}\bigg)^{j-n}\\& \nonumber\times \bigg(e^{-i\varphi_2}\sin\bigg(\frac{\theta_2}{2}\bigg)\bigg)^{j+n}\bigg) \exp{[-\frac{\rvert 2j-2\alpha+m+n
			\lvert^2}{2}+(\alpha-\alpha^{\ast})(m-n)]} \\& \nonumber\times \exp{[-\frac{\rvert 2j-2\beta-m-n
		\lvert^2}{2}+(\beta-\beta^{\ast})(n-m)]}.
	\end{align}
	In Fig.\ref{fig2}, we have chosen to illustrate the square of the Wigner function (\ref{nom28}) for the respective spin values under consideration, relative to selected positions or momenta (the other variables are set to zero). The two pairs $(\theta_1, \theta_2)$ and $(\varphi_1, \varphi_2)$ are set to $(\frac{\pi}{3}, \frac{\pi}{2})$ and $(0, 2\pi)$ respectively.

In these different plots, we observe that the symmetry admits non-zero maximum values with different peaks with respect to the spin value $j$. This shows that the symmetry is not broken (i.e.,  not zero) at all points in the study interval $[-10,10]$. More precisely, in Fig.$\ref{fig2}$ $\left(a\right)$ we see that for all considered values of $j$ there is a symmetry violation in the intervals $[-10,-1/2]$ and $[6,10]$ of $q_1$ values ($q_1$ is the real part of $\alpha$). However, for spin values $j$ greater than or equal to $5/2$, we observe a total symmetry violation in the entire interval $[-10,10]$ of $q_1$ values. On the other hand, if we consider the square of the Wigner function as a function of $q_2$, the real part of $\beta$, we observe total symmetry violation for values of $j$ greater than or equal to 4 (see Fig.$\ref{fig2}$ $\left(d\right)$). However, if we interpret the square of the Wigner function as a function of the imaginary parts of $\beta$ and $\alpha$ (see Fig.$\ref{fig2}$ $\left(b\right)$ and  Fig.$\ref{fig2}$ $\left(c\right)$), we observe a violation of the parity symmetry for values of $j$ greater than or equal to $1$.

  \section{The parity symmetry under Gaussian noise channel} \label{sec4}
	\subsection{Spin-1/2 cat states under Gaussian noise channel} 
	Gaussian noise channels occur naturally in quantum information. They can also be called classical noise channels or bosonic Gaussian channels, and have been extensively studied, with a focus on quantum and classical capacity \cite{Glauber1963,Amosov2020}, mathematically defined by random Weyl shifts in phase space, which appear in several physical scenarios, including optimal quantum cloning, quantum communication, and teleportation of continuous variable states. A unique class of Gaussian channels plays an important role in the transmission of information in quantum systems with continuous variables. In this section, we explore how the presence of a Gaussian noise channel affects the parity symmetry and asymmetry of spin-1/2 cat states. The action of the Gaussian noise channel $\Phi_s^1$, with noise parameter $s>0$, on the two-mode bosonic state $\hat{\rho}_{12}=\mid Cat, \frac{1}{2}\rangle\langle\frac{1}{2}, Cat\mid$ in Hilbert space $H=H_1 \otimes H_2$ is defined as \cite{lili2023, zhang2021} 
	\begin{equation}
	\begin{array}{ll}\label{nomm20} 
	\big(\Phi_s^1\otimes \mathcal{I}^2\big)\big(\rho_{12}\big)=\displaystyle\int_{\mathbb{C}} \big(D(z)\otimes \mathbf{1}_2\big)\rho_{12}\big(D(z)^+\otimes \mathbf{1}_2\big) d\mu_s(z),
	\end{array}
	\end{equation}
	where,  
	%for a , with the reduced state on mode 1 denoted as $\rho_1 = tr_{\rho_2}(\mid Cat, \frac{1}{2}\rangle\langle\frac{1}{2}, Cat\mid)$
	%where $\Phi_s^1$ is the Gaussian noise channel  acting on mode $1$ is defined as
	%\begin{equation}
	% \begin{array}{ll}
	% \Phi_s^1(\rho_1)=\int_{\mathbb{C}} D(z)tr_{\rho_2}(\mid Cat, \frac{1}{2}\rangle\langle\frac{1}{2}, Cat\mid)D(z)^+ d\mu_s(z),
	% \end{array}
	%\end{equation}
	$$ z=\frac{q+ip}{\sqrt{2}}\in \mathbb{C},\hspace{0.5cm} p,q\in \mathbb{R}, \hspace{0.5cm} d\mu_s(z)= e^{-|z|^2/s}\frac{d^2z}{\pi s},\hspace{0.5cm} s>0$$  
	and $\mathbf{1}_2$, $\mathcal{I}^2$ are, respectively,  the identity operators in the Hilbert space $H_2$ and the identity channel on mode $2$. 
    
     \begin{figure}[hbtp]
  			{{\begin{minipage}[b]{.2\linewidth}
  						\centering
  						\includegraphics[scale=0.20]{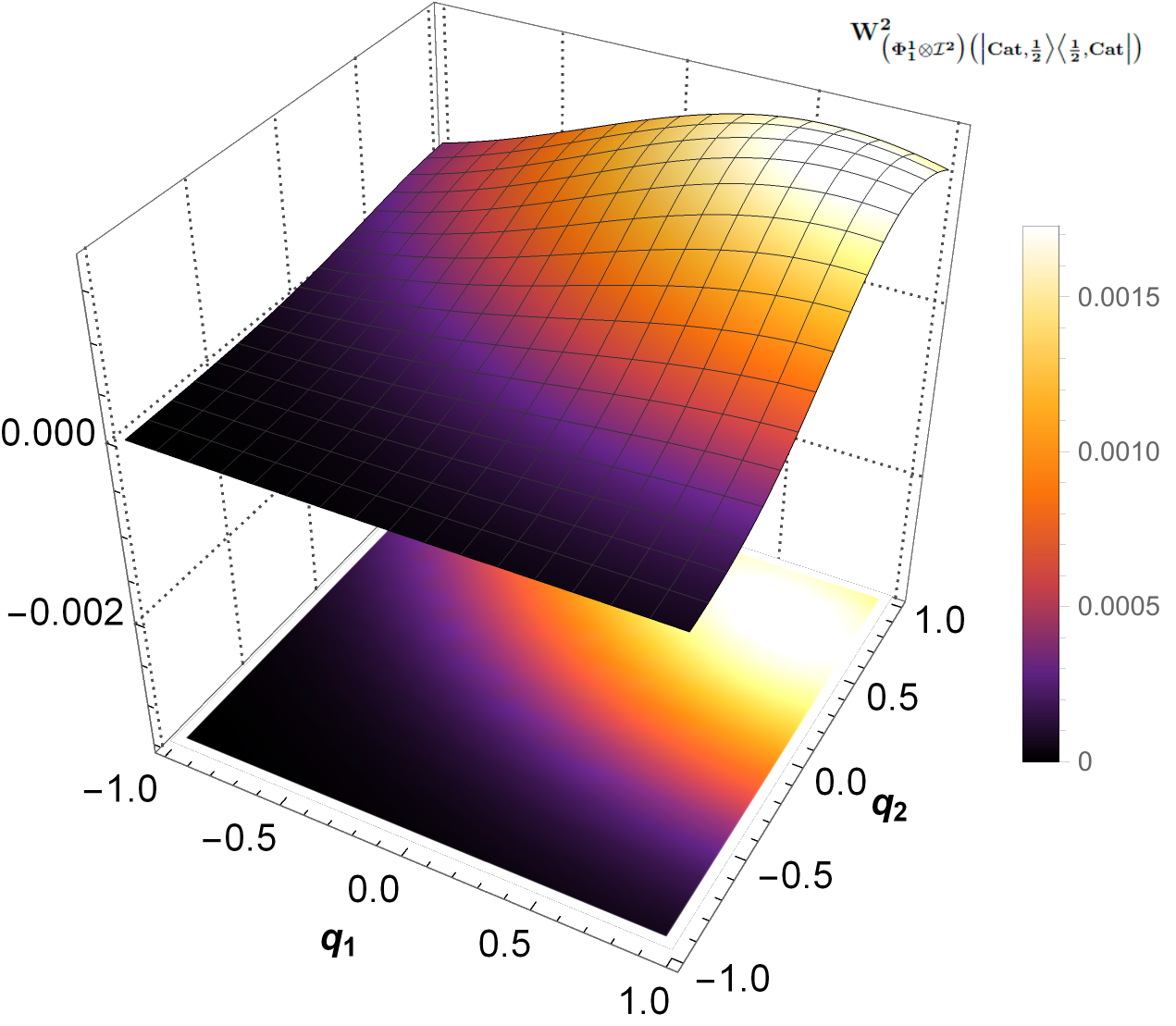} \vfill $\left(a\right)$
  					\end{minipage} \hfill
  					\begin{minipage}[b]{.2\linewidth}
  						\centering
  						\includegraphics[scale=0.20]{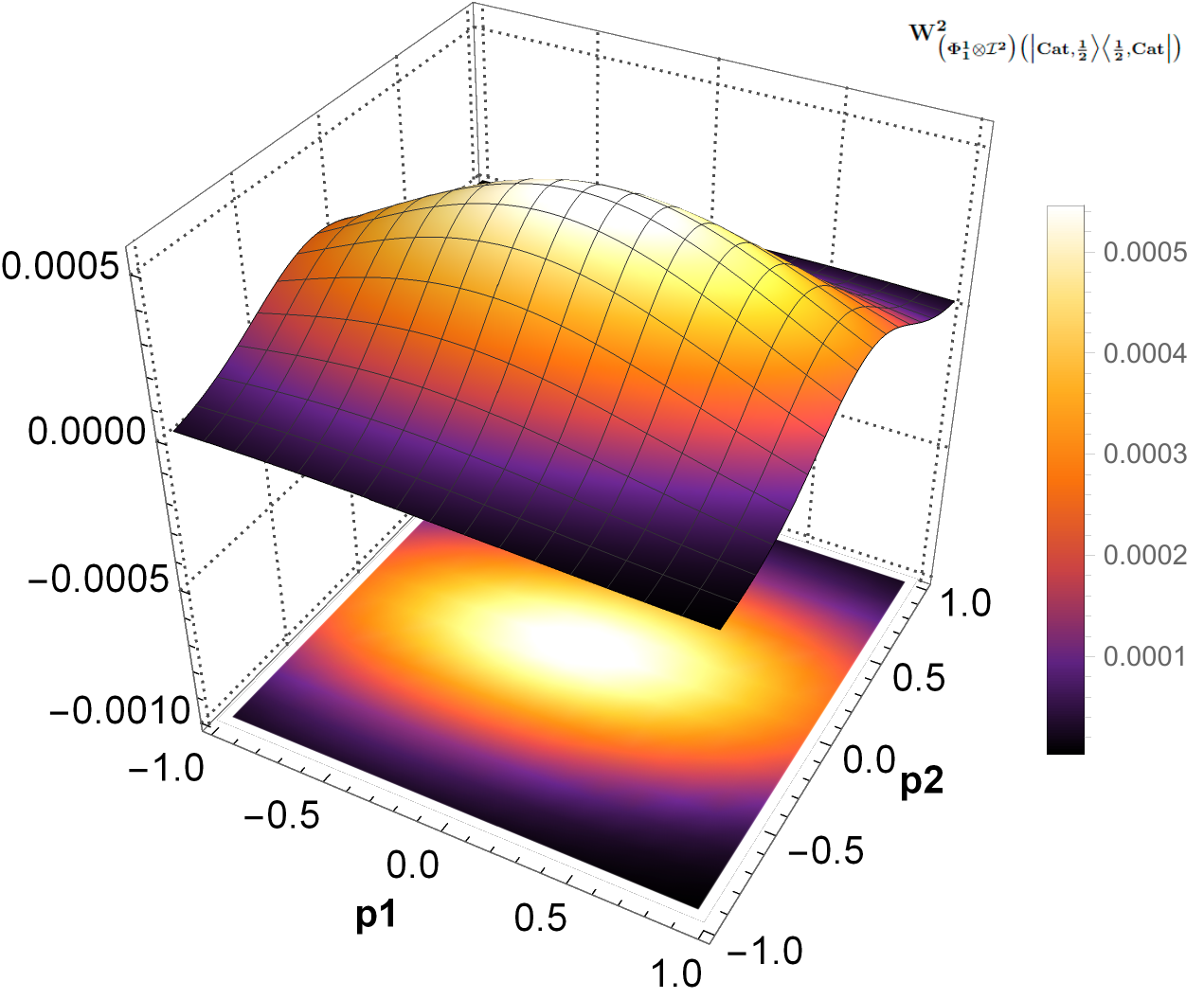} \vfill  $\left(b\right)$
 					\end{minipage} \hfill
  					\begin{minipage}[b]{.2\linewidth}
  						\centering
  						\includegraphics[scale=0.20]{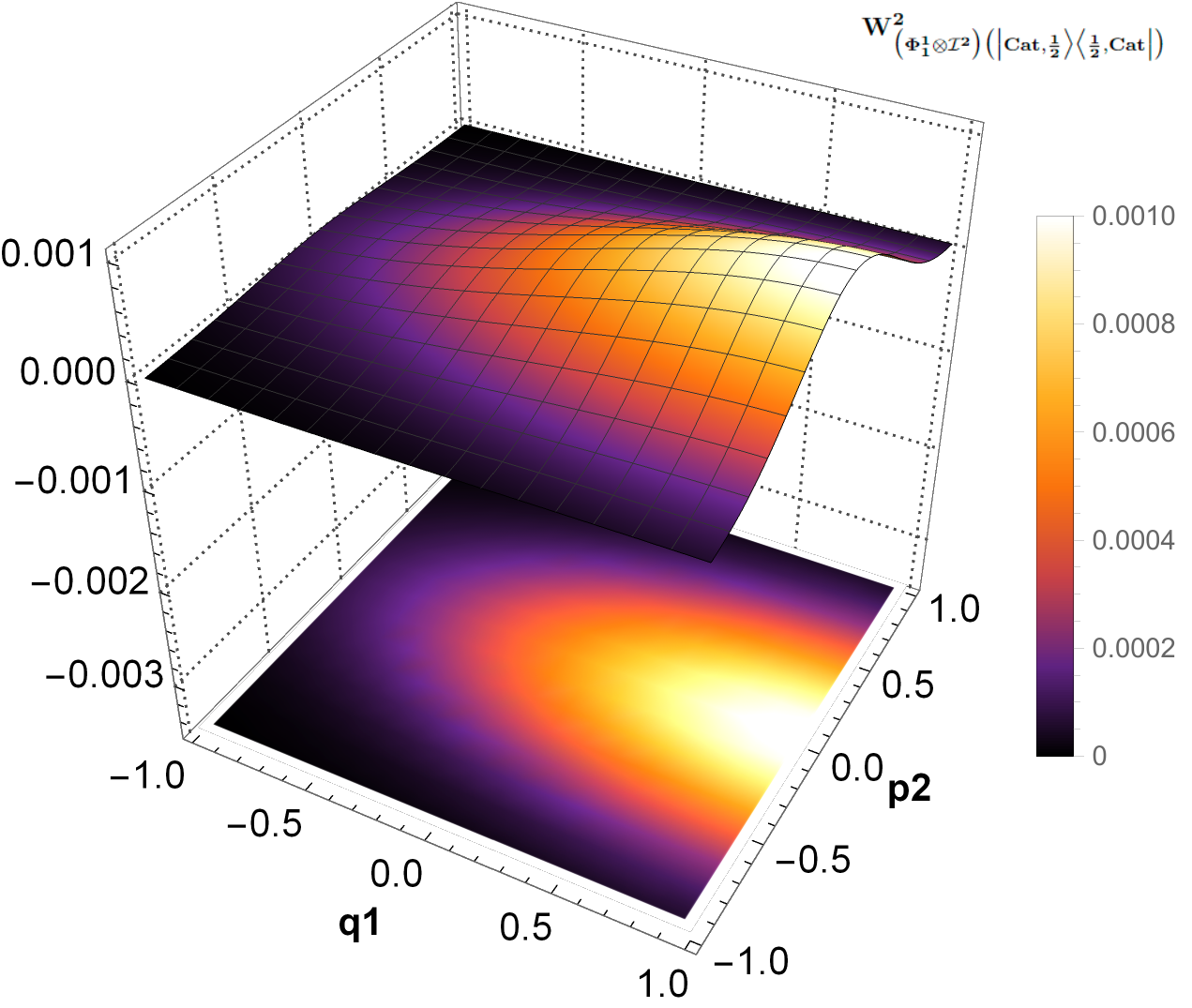} \vfill  $\left(c\right)$
  			\end{minipage} \hfill
     \begin{minipage}[b]{.2\linewidth}
  						\centering
  						\includegraphics[scale=0.20]{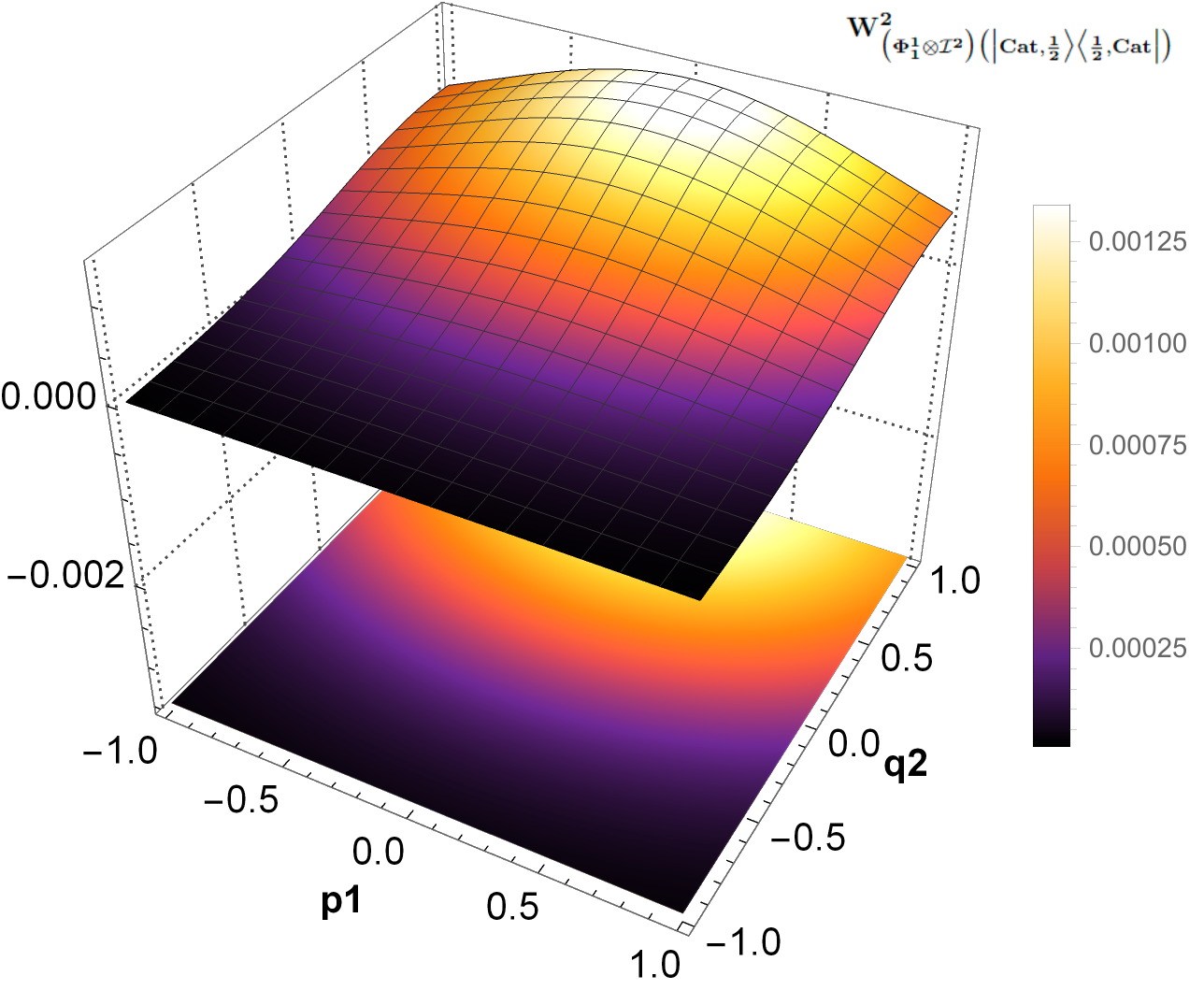} \vfill $\left(d\right)$
  				\end{minipage}}}
      {{\begin{minipage}[b]{.2\linewidth}
  						\centering
  						\includegraphics[scale=0.31]{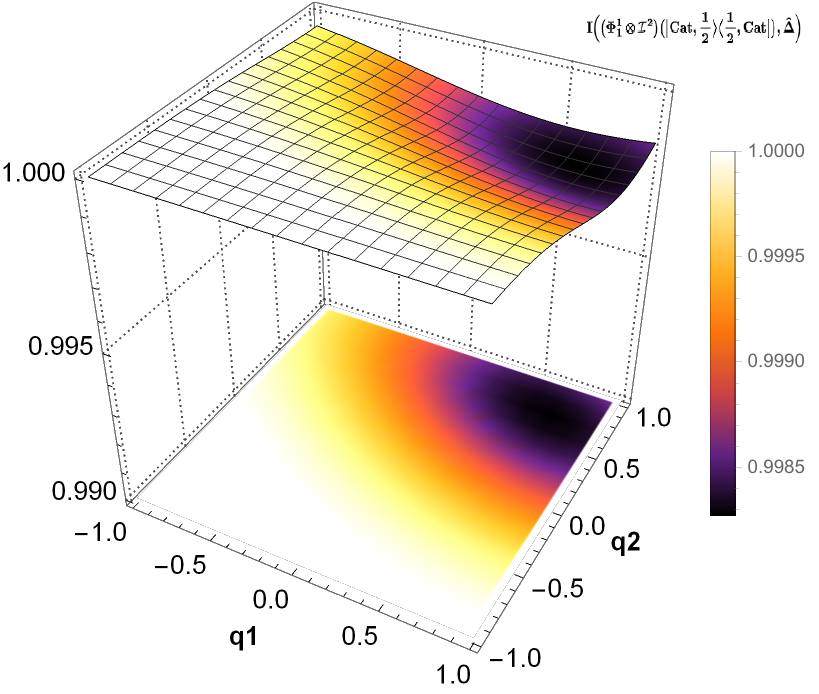} \vfill $\left(e\right)$
  					\end{minipage} \hfill
  					\begin{minipage}[b]{.2\linewidth}
  						\centering
  						\includegraphics[scale=0.31]{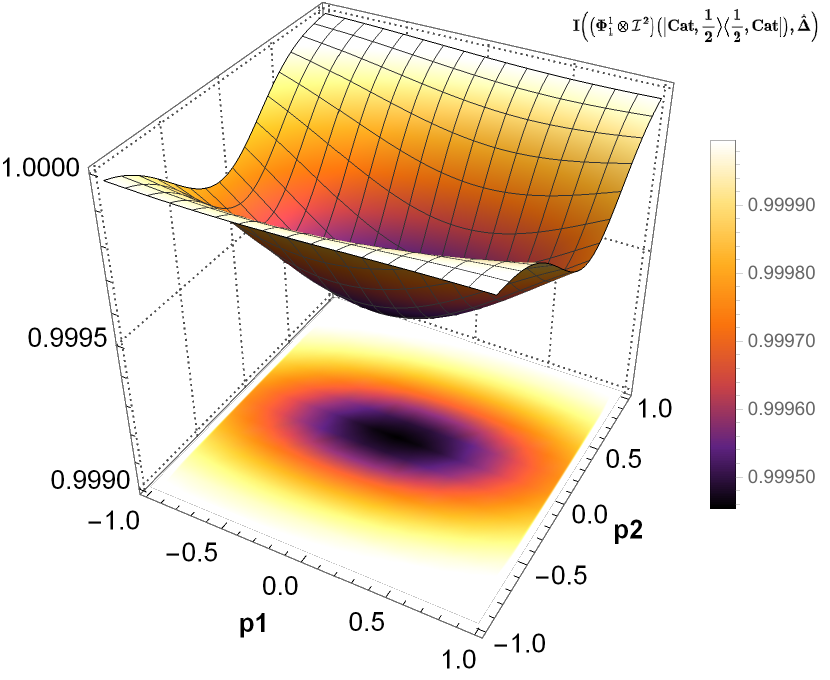} \vfill  $\left(f\right)$
 					\end{minipage} \hfill
  					\begin{minipage}[b]{.2\linewidth}
  						\centering
  						\includegraphics[scale=0.31]{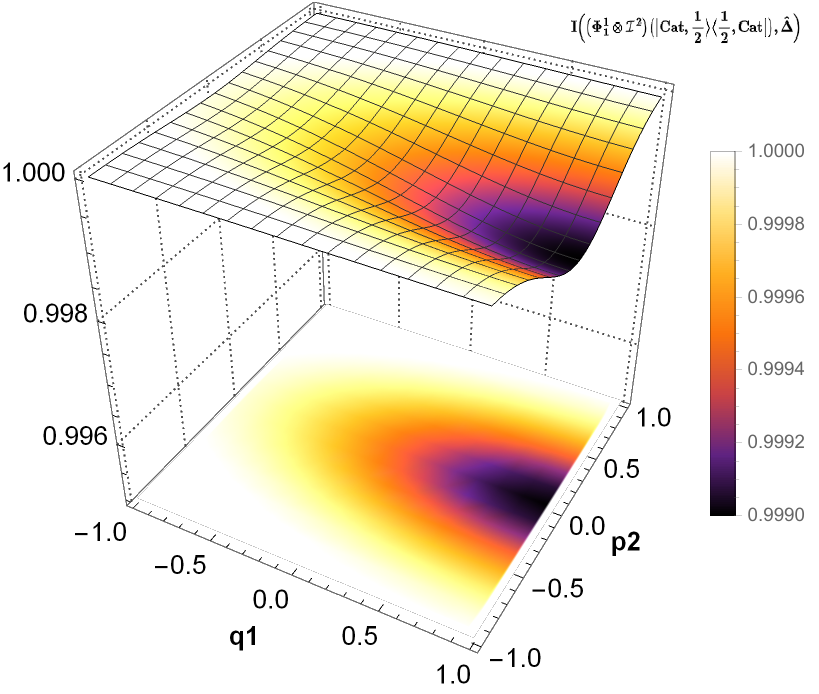} \vfill  $\left(g\right)$
  			\end{minipage} \hfill
     \begin{minipage}[b]{.2\linewidth}
  						\centering
  						\includegraphics[scale=0.31]{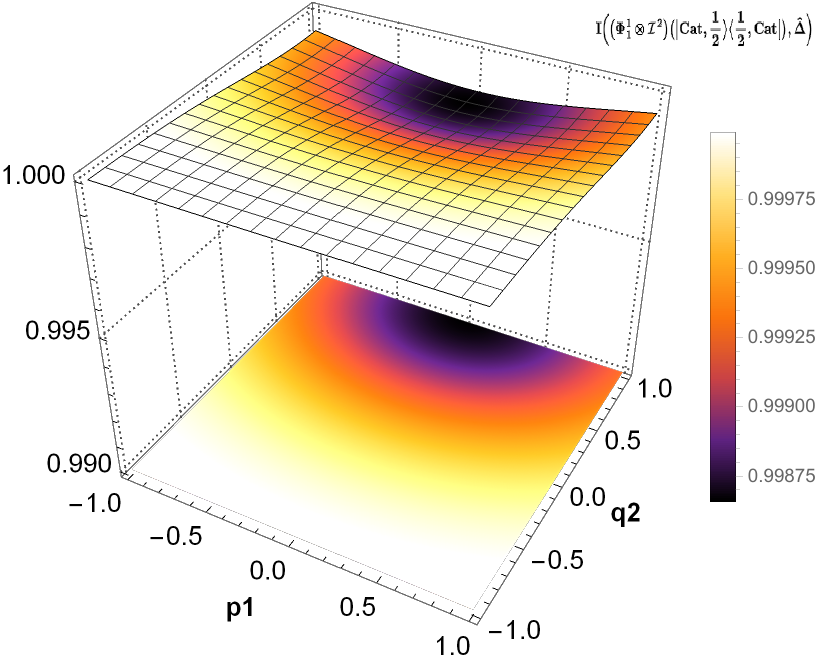} \vfill $\left(h\right)$
  				\end{minipage}}}
  \caption{The square of the Wigner function (\ref{the Gaussian noise channel}) and Wigner-Yanase skew information (\ref{nomm12}) of spin-1/2 cat state under the influence Gaussian noise channel in phase space: \textbf{(a)} square of the Wigner function and \textbf{(e)}  Wigner-Yanase skew information for $ p_1=p_2=0$, \textbf{(b)} square of the Wigner function and \textbf{(f)}  Wigner-Yanase skew information for $q_1=q_2=0$, \textbf{(c)} square of the Wigner function and \textbf{(g)}  Wigner-Yanase skew information for $p_1=q_2=0$, \textbf{(d)} square of the Wigner function and \textbf{(h)}  Wigner-Yanase skew information for $q_1=p_2=0$; with the noise parameter $s=1$.}
	    \label{fig3}
 	 	\end{figure}
With this effect, we derive the Wigner function expression as: 
       \begin{align}\label{the Gaussian noise channel}
	W_{\big(\Phi_s^1\otimes \mathcal{I}^2\big)\big(| Cat, \frac{1}{2}\rangle \langle\frac{1}{2}, Cat|\big)} \Big( q_1,q_2,p_1,p_2\Big)&:=W_{\big(\Phi_s^1\otimes \mathcal{I}^2\big)\big(| Cat, \frac{1}{2}\rangle \langle\frac{1}{2}, Cat|\big)} \Big( \alpha,\beta\Big)\\&\nonumber=  \bigg(\frac{\mathcal{N}}{\pi}\bigg)^2 \displaystyle\int_\mathbb{C}\bigg[ \bigg(\cos\bigg(\frac{\theta_1}{2}\bigg)+ \cos\bigg(\frac{\theta_2}{2}\bigg)\bigg)^2 e^{-2|z-\alpha|^2} e^{-2 \mid 1-\beta \mid^2} +\bigg(\cos \bigg(\frac{\theta_1}{2}\bigg)+ \cos\bigg(\frac{\theta_2}{2}\bigg)\bigg)\\& \nonumber
	\times \bigg(e^{-i \varphi_1} \sin\bigg(\frac{\theta_1}{2}\bigg)+e^{-i \varphi_2} \sin\bigg(\frac{\theta_2}{2}\bigg)\bigg) 
	e^{-\frac{1}{2} \mid 2z +1-2 \alpha\mid^2}e^{- \alpha+\alpha^\ast+ z-z^\ast} e^{-\frac{1}{2} \big(\mid 1-2 \beta\mid^2- 2\beta+2\beta^\ast\big)}\\& \nonumber  + \bigg(\cos \bigg(\frac{\theta_1}{2}\bigg)+\cos \bigg(\frac{\theta_2}{2}\bigg)\bigg)
	\bigg(e^{i \varphi_1} \sin\bigg(\frac{\theta_1}{2}\bigg)+e^{i \varphi_2}\sin\bigg(\frac{\theta_2}{2}\bigg)\bigg) 
	e^{-\frac{1}{2} \big(\mid 1-2 \beta\mid^2- 2\beta^\ast+2\beta\big)}  \\& \nonumber  \times e^{-\frac{1}{2}\mid 1- 2\alpha+2z\mid^2} e^{z^\ast-z - \alpha^\ast +\alpha} +\bigg(\sin^2\bigg(\frac{\theta_1}{2}\bigg)+2 \sin\bigg(\frac{\theta_1}{2}\bigg) \sin\bigg(\frac{\theta_2}{2}\bigg)\cos\bigg(\varphi_1-\varphi_2\bigg)
	\\& \nonumber +\sin^2\bigg(\frac{\theta_2}{2}\bigg)\bigg) 
	e^{-2\mid 1- \alpha+z\mid^2}  e^{-2\mid \beta \mid^2}\bigg] d\mu_s(z^2).
	\end{align}

In Fig.\ref{fig3}, we choose to show the plots of the square of the Wigner function and Wigner-Yanase skew information of spin-$1/2$ cat state under the influence of Gaussian noise channel as functions of selected positions and momenta (the other quadrature variables being put to zero) for the same values of $\theta_i$ and $\phi_i$ (i.e., $(\theta_1,\theta_2)=(\pi,0)$ and  $(\phi_1,\phi_2)=(0, 2\pi)$) and the noise parameter $s$ is fixed at $1$. The different plots show that the Gaussian noise channel preserves the evolution of parity symmetry (respectively, the parity asymmetry) in the previous section, but with a decrease in the degree of parity symmetry (respectively, with an increase in the degree of parity asymmetry). For example, concerning asymmetry, we see that the Gaussian noise channel forces the parity degree to its maximum value. On the other hand, for symmetry, such a channel forces the degree of parity to its minimum value.\par
\begin{figure}[hbtp]
 			{{\begin{minipage}[b]{.2\linewidth}
 						\centering
 						\includegraphics[scale=0.30]{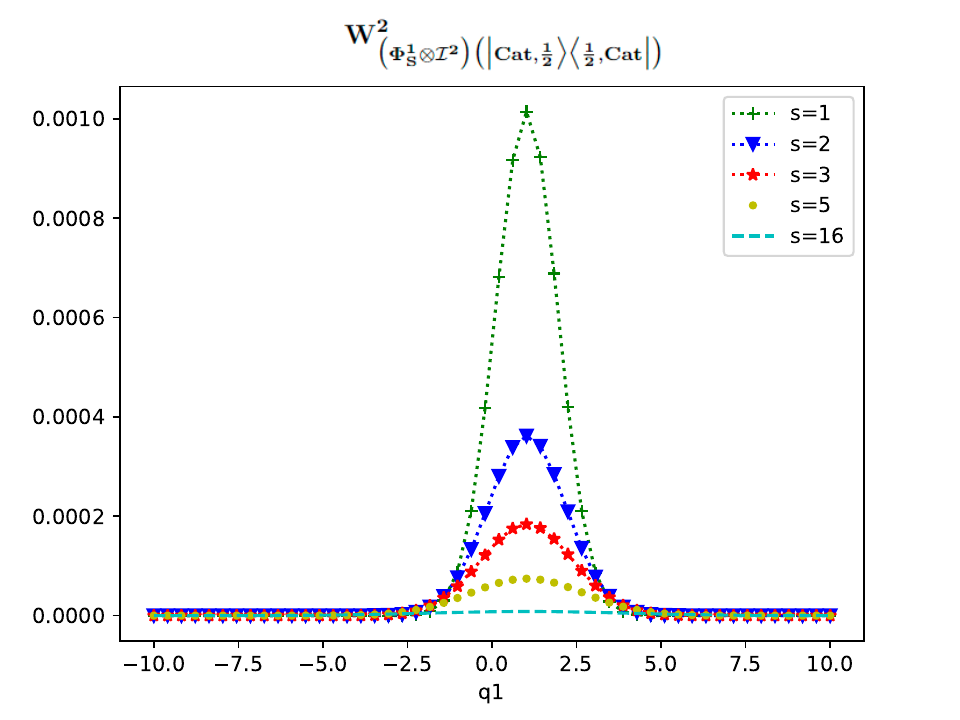} \vfill $\left(a\right)$
 					\end{minipage} \hfill
 					\begin{minipage}[b]{.2\linewidth}
 						\centering
 						\includegraphics[scale=0.30]{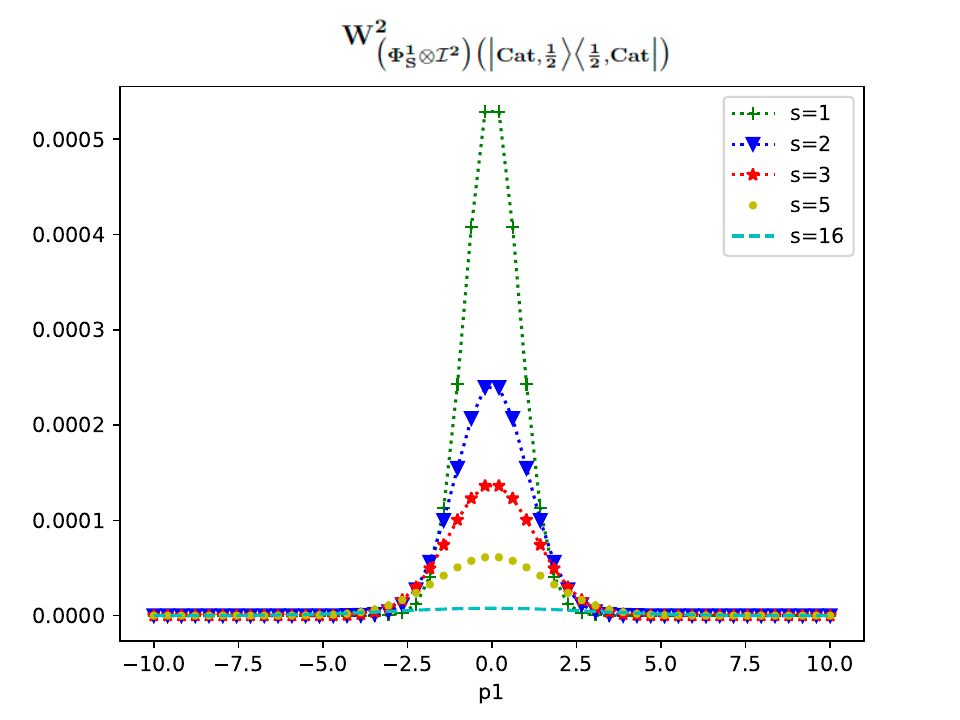} \vfill  $\left(b\right)$
 					\end{minipage} \hfill
 					\begin{minipage}[b]{.2\linewidth}
 						\centering
 						\includegraphics[scale=0.30]{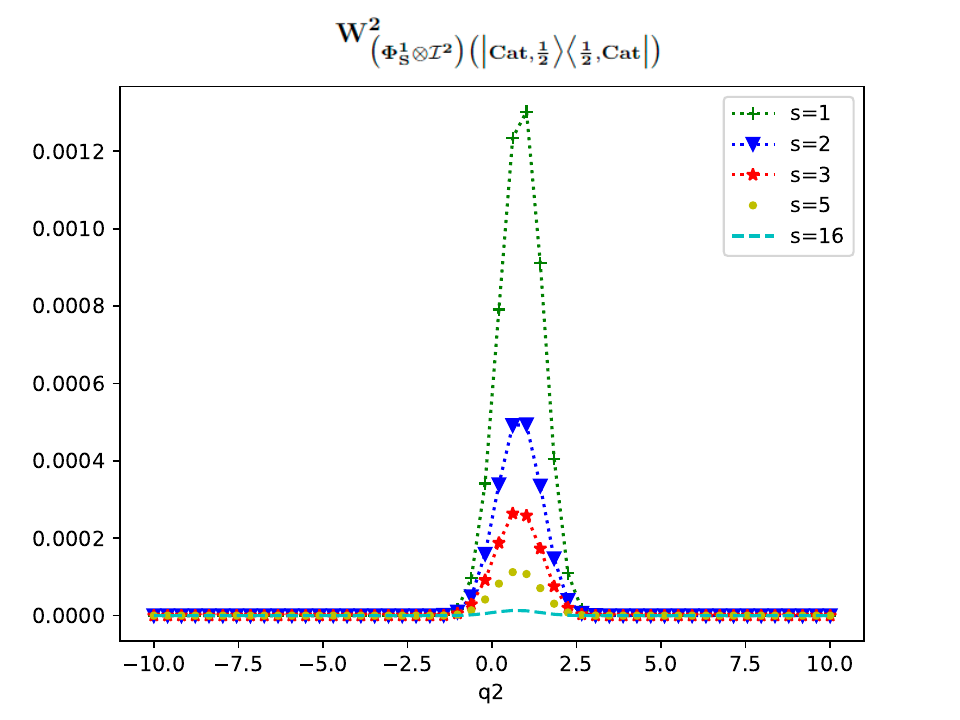} \vfill  $\left(c\right)$
 			\end{minipage} \hfill
    \begin{minipage}[b]{.2\linewidth}
 						\centering
 						\includegraphics[scale=0.30]{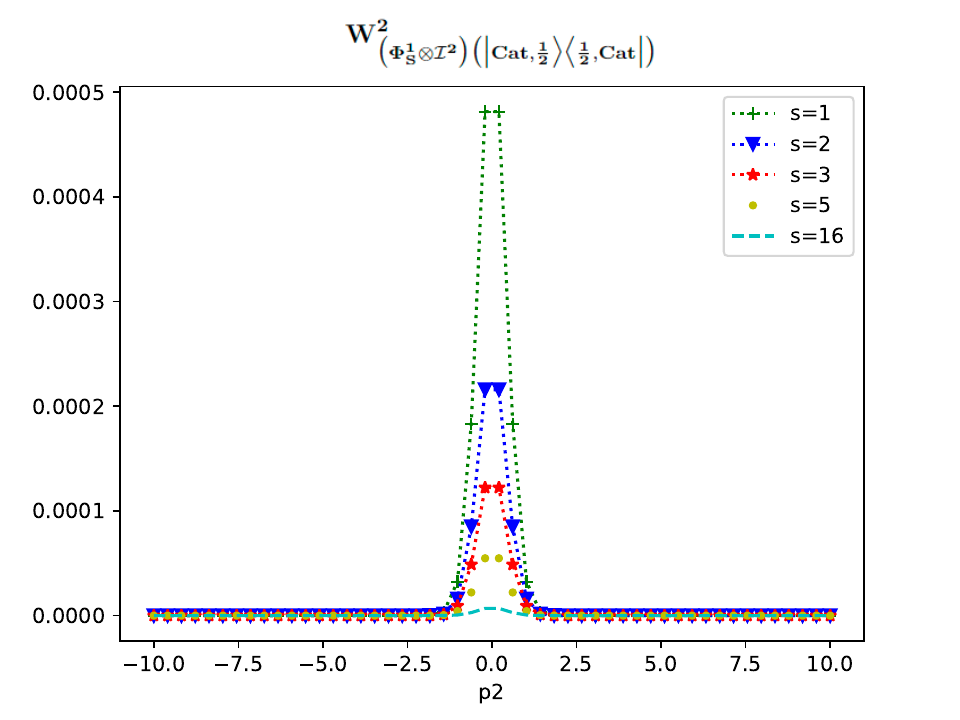} \vfill $\left(d\right)$
 					\end{minipage}}}
 \caption{The square of the Wigner function (\ref{the Gaussian noise channel}) of spin-1/2 cat state in terms of $q_1$, $p_1$, $q_2$ and  $p_2$  for different values of decoherence parameter $s$.}
		\label{fig4}
 		\end{figure}

	Since the noise channels depend on the parameter $s$, it's motivating to see how the parity varies with different channels. So we've plotted Fig.\ref{fig4}, which shows the variation of the  Wigner function squared in phase space.\par
	Fig.\ref{fig4} shows the dependence of the square of the Wigner function on different values of the parameter $s$ in phase space. Parity symmetry tends to reach its minimum value completely within the study interval as parameter $s$ takes on increasingly larger values. 

\subsection{Spin-$j$ cat states and Gaussian noise channel}
	Here we study how the presence of a Gaussian noise channel affects the parity symmetry of spin-j Cat states. The effect of the Gaussian noise channel $\Phi_s^1$, on the two-mode bosonic state $\mid Cat, j\rangle\langle j, Cat\mid$ in Hilbert space gives rise to the following Wigner function:
	\begin{align}\label{the Gaussian noise channelj}
	W_{\big(\Phi_s^1\otimes \mathcal{I}^2\big)\big(| Cat,j\rangle \langle j, Cat|\big)} \Big(q_1,q_2,p_1,p_2\Big)&:=W_{\big(\Phi_s^1\otimes \mathcal{I}^2\big)\big(| Cat,j\rangle \langle j, Cat|\big)} \Big( \alpha,\beta\Big)\\& \nonumber = \bigg[\displaystyle \int_\mathbb{C}  \exp{\Big[-\frac{\rvert 2j-2(\alpha+z)+m+n
			\lvert^2}{2}+(\alpha+z)-(\alpha^{\ast}+z^{\ast})(m-n)\Big]} d\mu_s(z^2)\bigg]\nonumber\\& \nonumber \times\bigg[ \left(\frac{\mathcal{N}_t}{\pi}\right)^2\sum_{m,n=-j}^j\left(\begin{array}{c}
	2 j \\
	j+m
	\end{array}\right)^{\frac{1}{2}} 
 \left(\begin{array}{c}
	2 j \\
	j+n 
	\end{array}\right)^{\frac{1}{2}}\bigg(\cos\bigg(\frac{\theta_1}{2}\bigg)^{j-m}\bigg(e^{i\varphi_1}\sin\bigg(\frac{\theta_1}{2}\bigg)\bigg)^{j+m}\\& \nonumber+\cos\bigg(\frac{\theta_2}{2}\bigg)^{j-m} \bigg(e^{i\varphi_2}\sin\bigg(\frac{\theta_2}{2}\bigg)\bigg)^{j+m}\bigg) \bigg(\cos\bigg(\frac{\theta_1}{2}\bigg)^{j-n}\bigg(e^{-i\varphi_1}\sin\bigg(\frac{\theta_1}{2}\bigg)\bigg)^{j+n}\\& \nonumber+\cos\bigg(\frac{\theta_2}{2}\bigg)^{j-n}\bigg(e^{-i\varphi_2}\sin\bigg(\frac{\theta_2}{2}\bigg)\bigg)^{j+n}\bigg) \exp{\Big[(\beta-\beta^{\ast})(n-m)-\frac{\rvert 2j-2\beta-m-n
			\lvert^2}{2}\Big]} \bigg].
\end{align}
	
 \begin{figure}[hbtp]
 			{{\begin{minipage}[b]{.2\linewidth}
 						\centering
 						\includegraphics[scale=0.30]{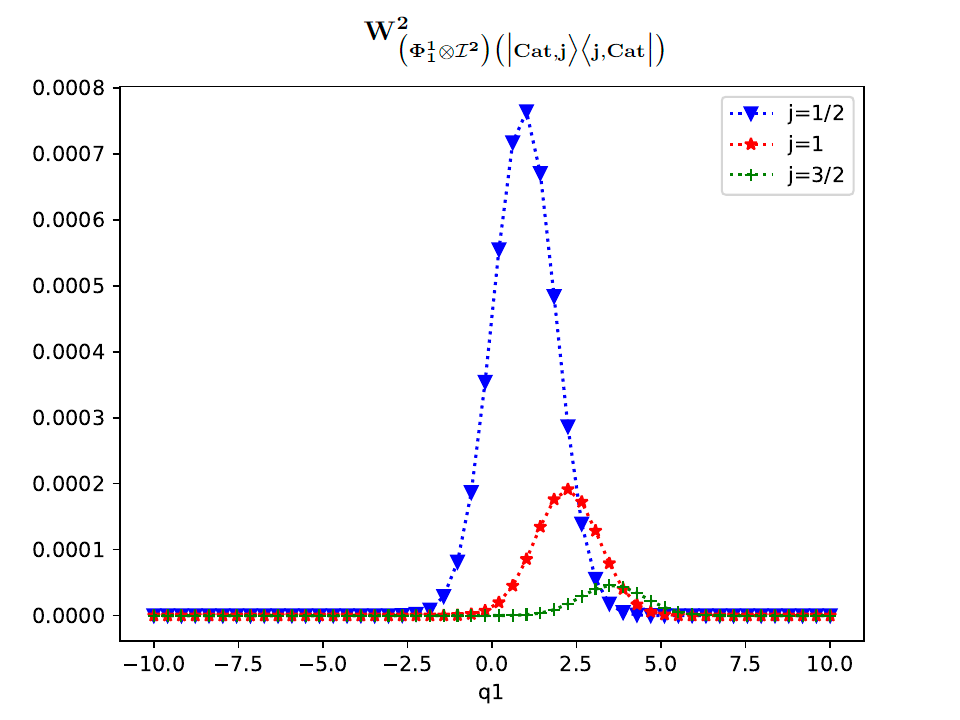} \vfill $\left(a\right)$
 					\end{minipage} \hfill
 					\begin{minipage}[b]{.2\linewidth}
 						\centering
 						\includegraphics[scale=0.30]{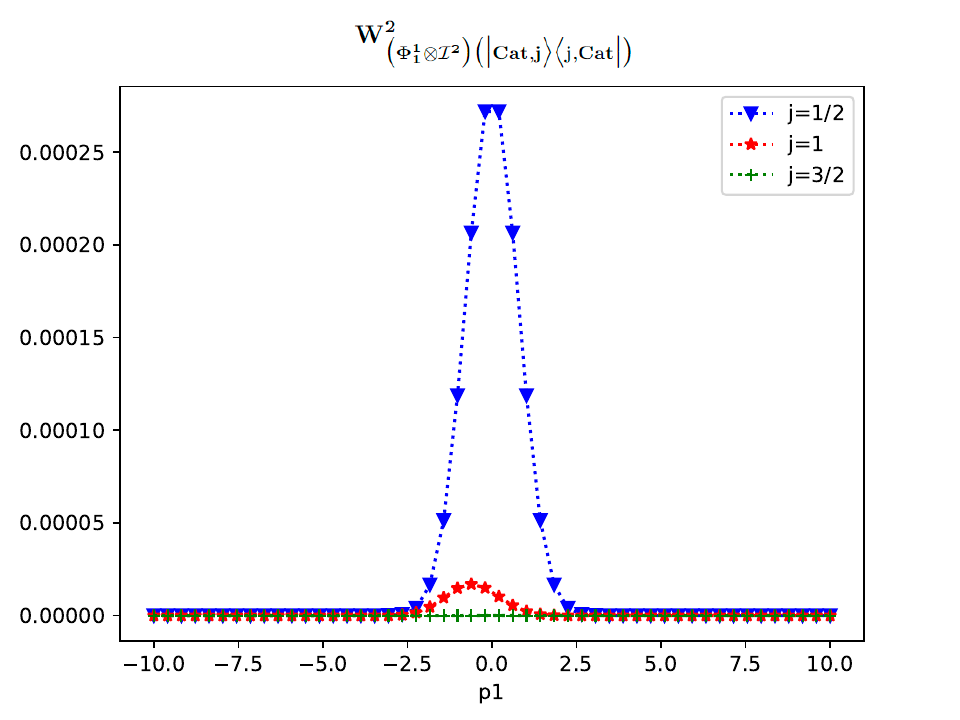} \vfill  $\left(b\right)$
 					\end{minipage} \hfill
 					\begin{minipage}[b]{.2\linewidth}
 						\centering
 						\includegraphics[scale=0.30]{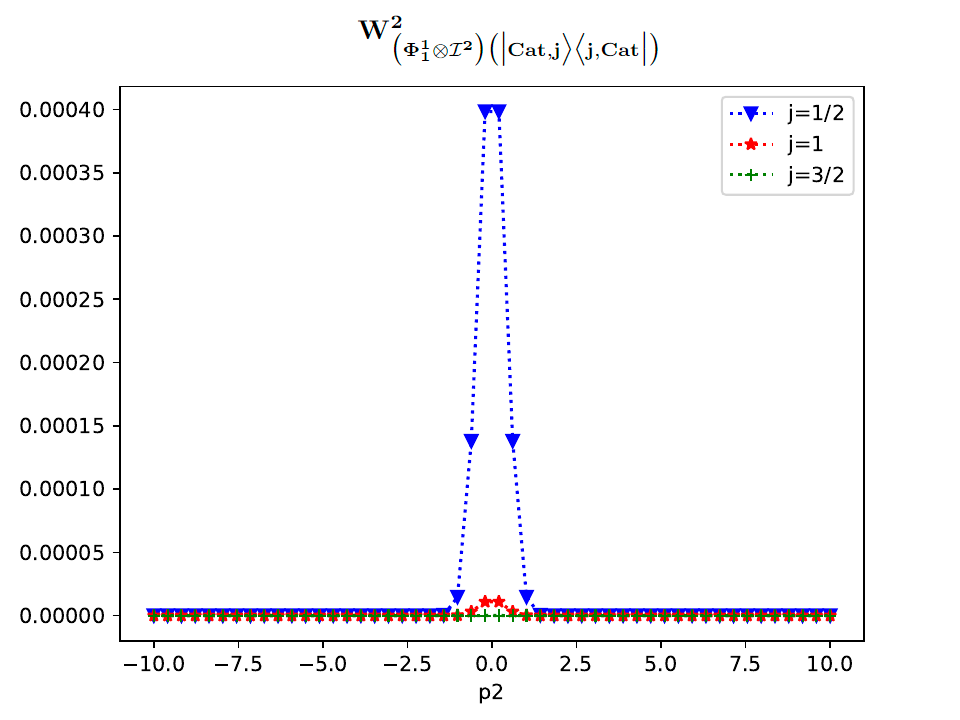} \vfill  $\left(c\right)$
 			\end{minipage} \hfill
    \begin{minipage}[b]{.2\linewidth}
 						\centering
 						\includegraphics[scale=0.30]{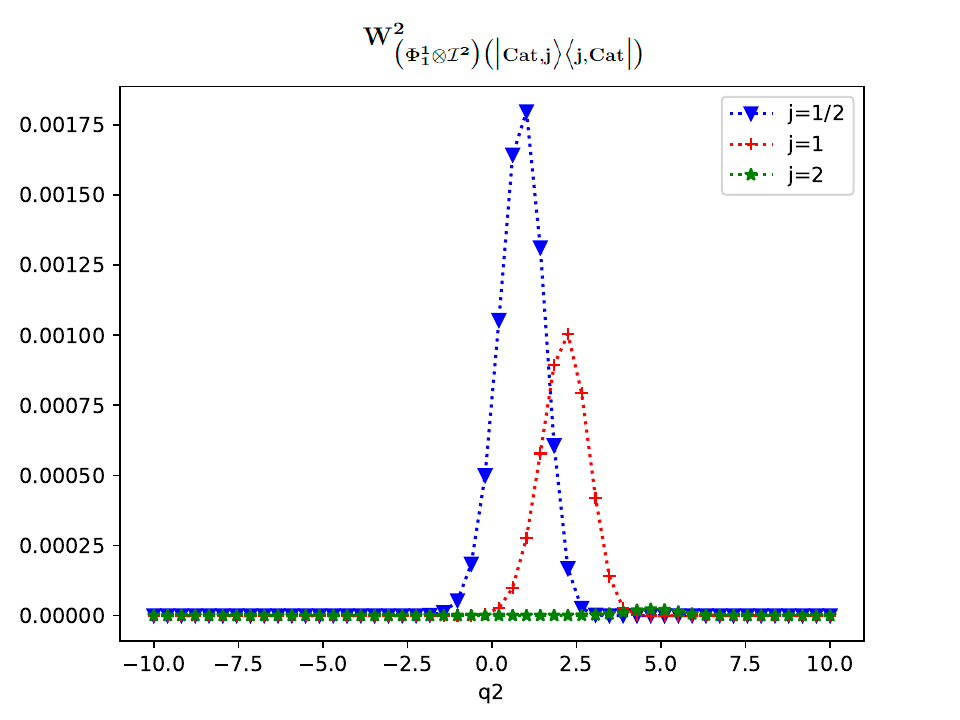} \vfill $\left(d\right)$
 					\end{minipage}}}
      {{\begin{minipage}[b]{.2\linewidth}
 						\centering
 						\includegraphics[scale=0.30]{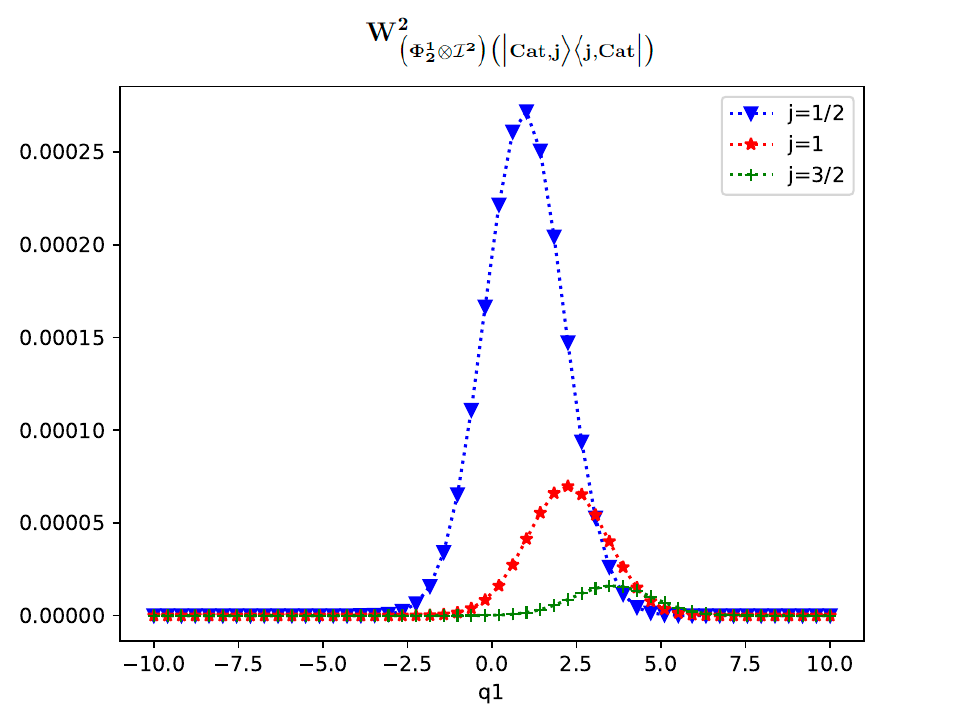} \vfill $\left(e\right)$
 					\end{minipage} \hfill \begin{minipage}[b]{.2\linewidth}
 						\centering
 						\includegraphics[scale=0.30]{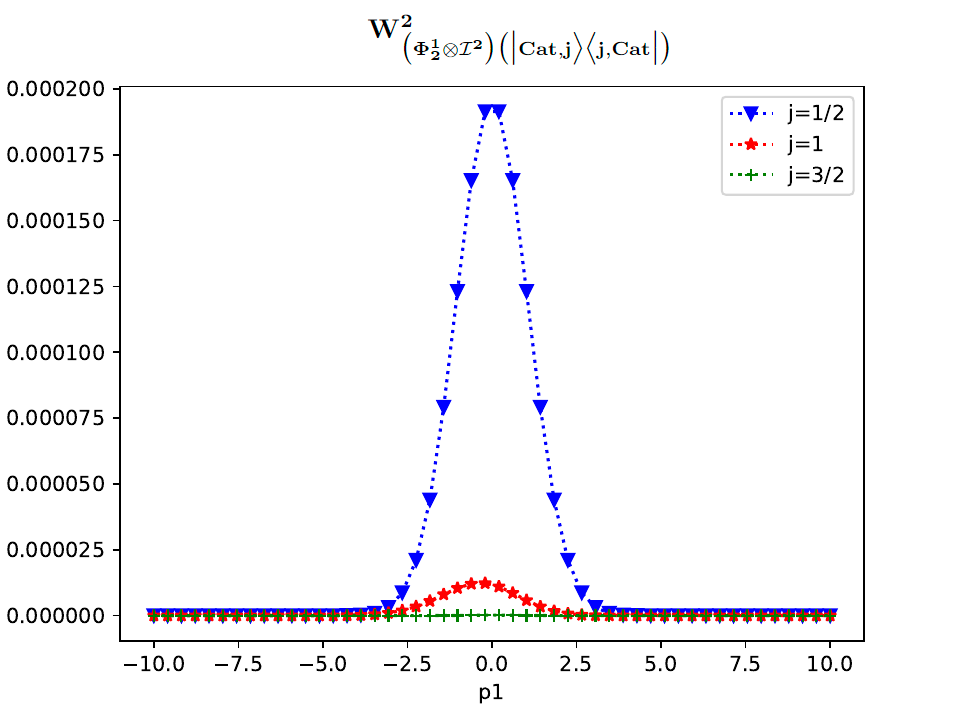} \vfill  $\left(f\right)$
 					\end{minipage} \hfill \begin{minipage}[b]{.2\linewidth}
 						\centering
 						\includegraphics[scale=0.30]{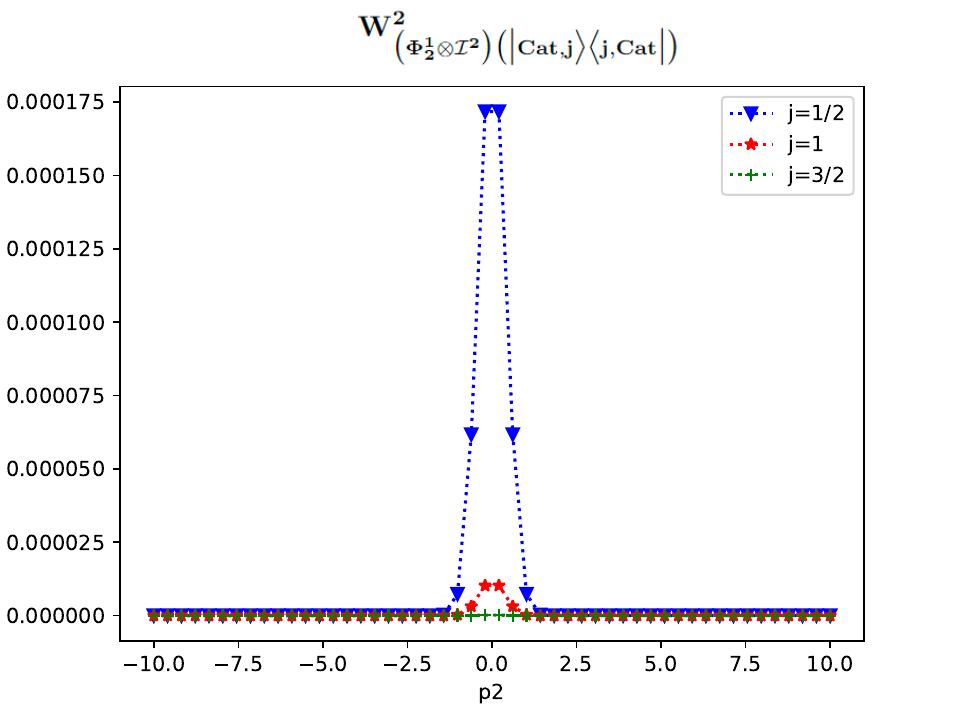} \vfill  $\left(g\right)$
 			\end{minipage} \hfill
    \begin{minipage}[b]{.2\linewidth}
 						\centering
 						\includegraphics[scale=0.30]{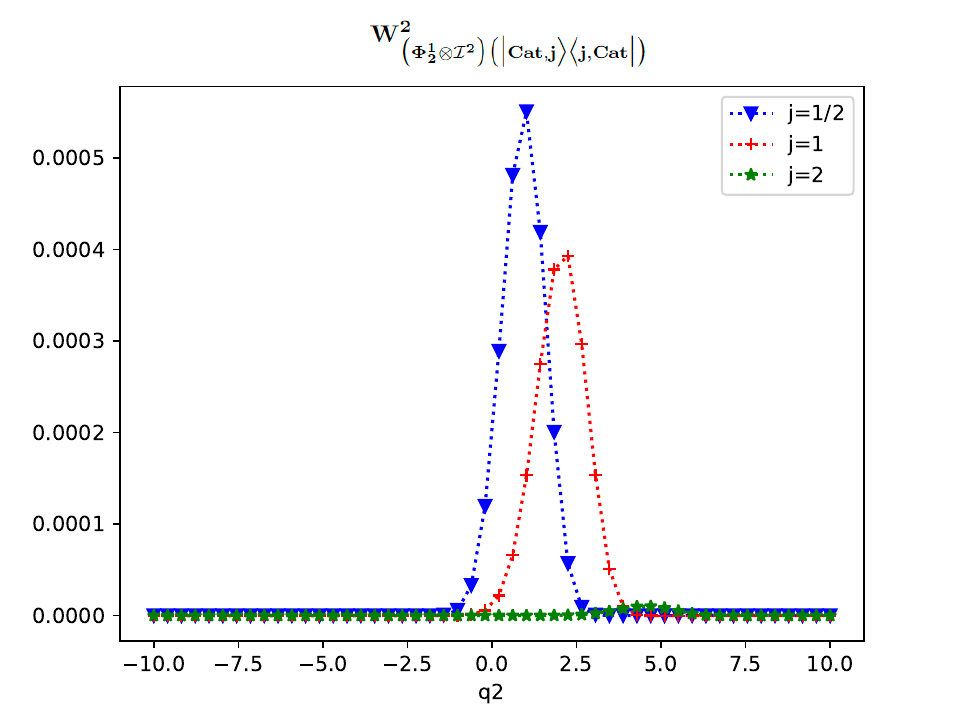} \vfill $\left(h\right)$
 					\end{minipage}}}
\caption{The square of the Wigner function (\ref{the Gaussian noise channelj}) of spin-$j$ cat state under the influence of Gaussian noise channel in terms of $q_{i}$ and $p_{i}$ (with $i=1,2$) for different values of spin $j$. The noise parameter $s=1$ for Panel ($a$) to Panel ($d$), and $s=2$ for Panel ($e$) to Panel ($h$).}\label{fig5}
 \end{figure}

To observe the effect of noise channels on our system, we plotted the square of the Wigner function in phase space for different noise channel parameters; $s=1$ for panel Fig.\ref{fig5}$\left(a\right)$, to panel Fig.\ref{fig5}$\left(d\right)$ and $s=2$ for panel Fig.\ref{fig5}$\left(e\right)$ to panel Fig.\ref{fig5}$\left(h\right)$. As shown in Fig.\ref{fig5}, the effect of these channels preserves the evolution of parity symmetry for any value of spin $j$, but the values of the square of the Wigner function decrease, i.e. the degree of parity symmetry decreases. Thus, we can generalize the effect of the noise channels on the spin-$\frac{1}{2}$ state for any spin value $j$. \par 
The phenomenon of parity symmetry breaking can be attributed to a number of factors, including quantum interference, which generates asymmetric distributions; non-classical correlations between position and momentum; and decoherence effects, which are introduced as a result of interaction with the environment. These factors can also give rise to asymmetries in the Wigner function. From a mathematical standpoint, the Wigner function, a mathematical tool utilized in quantum physics, exhibits asymmetry due to its intrinsic properties and its interconnection with the non-classical behavior observed in quantum systems. In contrast to classical probability distributions, the Wigner function is capable of containing negative values, which reflect the phenomenon of quantum interference and result in the emergence of asymmetric distributions. On the other hand, the Wigner function is defined as a partial Fourier transform of the density matrix. Consequently, when the latter violates parity symmetry, by containing antisymmetric terms for example, both the former and its square will also exhibit asymmetry, which is evident in both the real and imaginary parts of the function. To sum up, the asymmetry observed in the Wigner function results from the combination of symmetry-breaking interactions in physical systems and the mathematical properties of the density matrix and its Wigner representation.
 
\section{Closing remarks}

 The Wigner function is crucial for understanding and analyzing the quantitative properties of systems in phase space. It is used to characterize the states of quantum systems such as coherent states, and composite states, and plays an essential role in measuring resources in quantum metrology and investigating the nonclassicality of states. In this study, we have employed the conservation relations between the Wigner function and the Wigner-Yanase skew information, which quantifies the degree to which a quantum state lacks symmetry with respect to an observable, as a framework for measuring and studying parity symmetry for spin-$j$ cat coherent states in terms of the Wigner function.\par

As conclusion, we have quantified and studied the behavior of the parity symmetry and asymmetry for the coherent spin state superposition based on the symmetry-asymmetry manifestations, where the symmetry part is represented by the Wigner function squared and the asymmetry part is represented by the Wigner-Yanase skew information. As a consequence, parity symmetry violation for spin-$1/2$ cat states is observed in remote regions where $q_{2}>0$ and $q_{1}>0$ (Fig.\ref{fig1} $\left(a\right)$,$ \left(e\right)$), when $-1\leq p_{2}\leq 0$ and $0\leq p_{1}\leq 1$, as well as when $p_2$ and $p_1$ are close to $1$ or $-1$. Furthermore, the phenomenon is observed when $-1 \leq  p_{1} \leq  -1/2$ and $1/2 \leq  p_{2} \leq  1$ (Fig.\ref{fig1}  $\left(b\right)$, $\left(f\right)$), as well as for $p_{2} = -1$ and $p_{2} = 1$, or when $-1 \leq  q_{1} \leq  -1/2$ (Fig.\ref{fig1} $\left(c\right)$, $\left(g\right)$). Similarly, in regions where $p_{1}=-1$ and $p_1=1$, or $-1\leq q_{2}\leq -1/2$ (Fig.\ref{fig1}  $\left(d\right)$, $\left(h\right)$), parity symmetry is violated. Conversely, in regions that are far from these conditions, parity symmetry is preserved. Moreover, in order to ascertain whether the parity symmetry or asymmetry of these states is affected by decoherence, we integrated our state into a Gaussian noise channel. It was observed that the parity symmetry was violated to an increasing extent as the $s$ parameter of the channel was increased. This indicates that the spin-$1/2$ cat state became totally asymmetric with respect to the diplaced parity operator (the kernel operator). Additionally, the impact of the $j$-value was examined, revealing that parity symmetry is entirely violated at specific $j$-values. This indicates that the spin-$j$ cat state becomes asymmetric with respect to the diplaced parity operator.\par
 
In summary, the parity asymmetry of the superposition spin coherent states, relative to the kernel operator or the two-mode displaced parity operator, is influenced by the environment. When considering Gaussian noise channel as the environment, the parity asymmetry is maximized for larger decoherence parameter $s$. Moreover, this behavior varies with the spin value $j$. Consequently, the parity symmetry of the superposition of coherent spin states is related to both decoherence and the spin value.\\

{\bf Declaration of competing interest:}\par
The authors declare that they have no known competing financial interests or personal relationships that could have appeared to influence the work reported in this paper.\\

{\bf Data availability:}\par
No data was used for the research described in the article.
%\begin{thebibliography}{1}

\end{document}